\documentclass[acmlarge]{acmart}

\setcopyright{acmcopyright}
\copyrightyear{2021}
\acmYear{2021}
\acmDOI{10.1145/1122445.1122456}

\usepackage{makecell}
\usepackage{subfig}
\usepackage{wrapfig}
\usepackage{rotating}

\newcommand\scalemath[2]{\scalebox{#1}{\mbox{\ensuremath{\displaystyle #2}}}}

\usepackage{algorithm}
\usepackage{algpseudocode}

\newcommand{\cctpbsl}{
\begin{bmatrix}
  1 & 0 & 0 \\
  0 & 1 & 0 \\
  0 & 0 & 1 
\end{bmatrix}
}

\newcommand{\cczpbs}{
\begin{bmatrix}
  1 & 0 & 0 & -1 &  \phantom{-}1 &  \phantom{-}1 & -1\\
  0 & 1 & 0 & \phantom{-}1 & -1 &  \phantom{-}1 & -1\\
  0 & 0 & 1 &  \phantom{-}1 &  \phantom{-}1 & -1 & -1 
\end{bmatrix}
}

\newcommand{\bcczpbs}{
\begin{bmatrix}
  2 & 0 & 0 & -1 &  \phantom{-}1 &  \phantom{-}1 & -1\\
  0 & 2 & 0 & \phantom{-}1 & -1 &  \phantom{-}1 & -1\\
  0 & 0 & 2 &  \phantom{-}1 &  \phantom{-}1 & -1 & -1 
\end{bmatrix}
}
\newcommand{\bccrodecbs}{
\begin{bmatrix}
  -1 &  \phantom{-}1 &  \phantom{-}1 & -1\\
  \phantom{-}1 & -1 &  \phantom{-}1 & -1\\
  \phantom{-}1 &  \phantom{-}1 & -1 & -1 
\end{bmatrix}
}

\newcommand{\fccsix}{
\begin{bmatrix}
1 &           -1 & 1 & \phantom{-}1 & 0 & \phantom{-}0 \\
1 & \phantom{-}1 & 0 & \phantom{-}0 & 1 & -1 \\
0 & \phantom{-}0 & 1 & -1           & 1 & \phantom{-}1
\end{bmatrix}
}

\title{Automatic Generation of Interpolants for Lattice Samplings: Part I --- Theory and Analysis}
\author{Joshua Horacsek}
\email{joshua.horacsek@ucalgary.ca}
\affiliation{%
  \institution{University of Calgary}
  \streetaddress{2500 University Dr. NW}
  \city{Calgary}
  \state{Alberta}
  \postcode{T2N 1N4}
}
\author{Usman Alim}
\email{ualim@ucalgary.ca}
\affiliation{%
  \institution{University of Calgary}
  \streetaddress{2500 University Dr. NW}
  \city{Calgary}
  \state{Alberta}
  \postcode{T2N 1N4}
}
\begin{abstract}
Interpolation is a fundamental technique in scientific computing and is at the heart of many scientific visualization techniques. There is usually a trade-off between the approximation capabilities of an interpolation scheme and its evaluation efficiency. For many  applications, it is important for a user to be able to navigate their data in real time. In practice, the evaluation efficiency (or speed) outweighs any incremental improvements in reconstruction fidelity. In this two-part work, we first analyze from a general standpoint the use of compact piece-wise polynomial basis functions to efficiently interpolate data that is sampled on a lattice. In the sequel, we detail how we generate efficient implementations via automatic code generation on both CPU and GPU architectures. Specifically, in this paper, we propose a general framework that can produce a fast evaluation scheme by analyzing the algebro-geometric structure of the convolution sum for a given lattice and basis function combination. We demonstrate the utility and generality of our framework by providing fast implementations of various box splines on the Body Centered and Face Centered Cubic lattices, as well as some non-separable box splines on the Cartesian lattice. We also provide fast implementations for certain Voronoi splines that have not yet appeared in the literature. Finally, we demonstrate that this framework may also be used for non-Cartesian lattices in 4D.
\end{abstract}



\begin{document}



\maketitle


\section{Introduction}
Given a set of discrete data points that reside on a lattice, one of the key fundamental operations that one can perform on those data is {\em interpolation}. This appears in many different contexts in science and engineering: one may use interpolation to fill in data between points in time-series data, to fill in missing data between pixels in an image, or missing data between voxels in a volumetric image. In fact, many scientific visualization algorithms require a continuous representation of a signal as an input. 

As simple as it may seem, interpolation of lattice sampled  data can quickly become complicated. To many, interpolation is synonymous with linear interpolation, and if we wish to interpolate in higher dimensions, we can simply linearly interpolate in each dimension separately (i.e. a tensor product extension). However, there is much more freedom available that we may take advantage of when we move past one dimension. For example, there are many alternative interpolants (or \emph{basis functions}) we may choose instead of a simple linear interpolant, many with higher accuracy and/or higher smoothness than a simple linear interpolant. Not only this, but we may have more freedom in terms of where samples are placed when the dimension of the space is $\ge 2$. We say `may' because for some problems, one is forced to use a specific lattice structure - typically the Cartesian lattice, whereas other problems allow lattice choice as a degree of freedom.

In this work, we build a framework for interpolation on lattices, and show how to do it {\em fast}. Our work originated in scientific visualization, in particular working with volumetric data, but the observations we make are generally applicable to other domains as well as higher dimensions. We will mainly keep to the 3-dimensional case, but we will note how to generalize to higher dimensions when it is necessary.

Returning to the context of scientific visualization, interpolation speed is important in practice because it facilitates interactivity which, in turn, allows for fast iteration over data-sets --- the ability to quickly (visually) explore data is one of the key strengths of visualization. But more often than not, in practice, interpolation boils down to ``use trilinear interpolation'', and it is clear why: trilinear interpolation is fast, it is implemented in hardware for all modern graphics processing units; it is straightforward to use, it inlvolves a single texture fetch instruction for modern GPUs; and it looks ``good enough'' in practice. However, linear interpolation has a number of shortcomings. Perhaps the most obvious is smoothness. Trilinear interpolation is based on a piece-wise linear univariate  function extended via tensor product along all three dimensions; this ensures a continuous approximation but introduces discontinuities in the first and higher order derivatives. While this could be mitigated by using filtering methods, there is additional memory overhead  associated with such methods~\cite{alim2010gradient}. Linear interpolation also introduces visual artifacts when interpolating at lower resolutions. This is again related to the smoothness of the trilinear interpolant, but also has deeper roots in sampling and approximation theory. In short, the geometry and support of the chosen interpolant dictate how a reconstruction attenuates high-frequency content in the Fourier domain. This gives rise to different types of artifacts in data reconstruction~\cite{retailor}. Thus, there is value in exploring the design space of interpolants with the goal of minimizing error; in practice, one also needs to account for the computational costs associated with different interpolants. 

\begin{figure*}[ht!]
    \centering
    \includegraphics[scale=0.135]{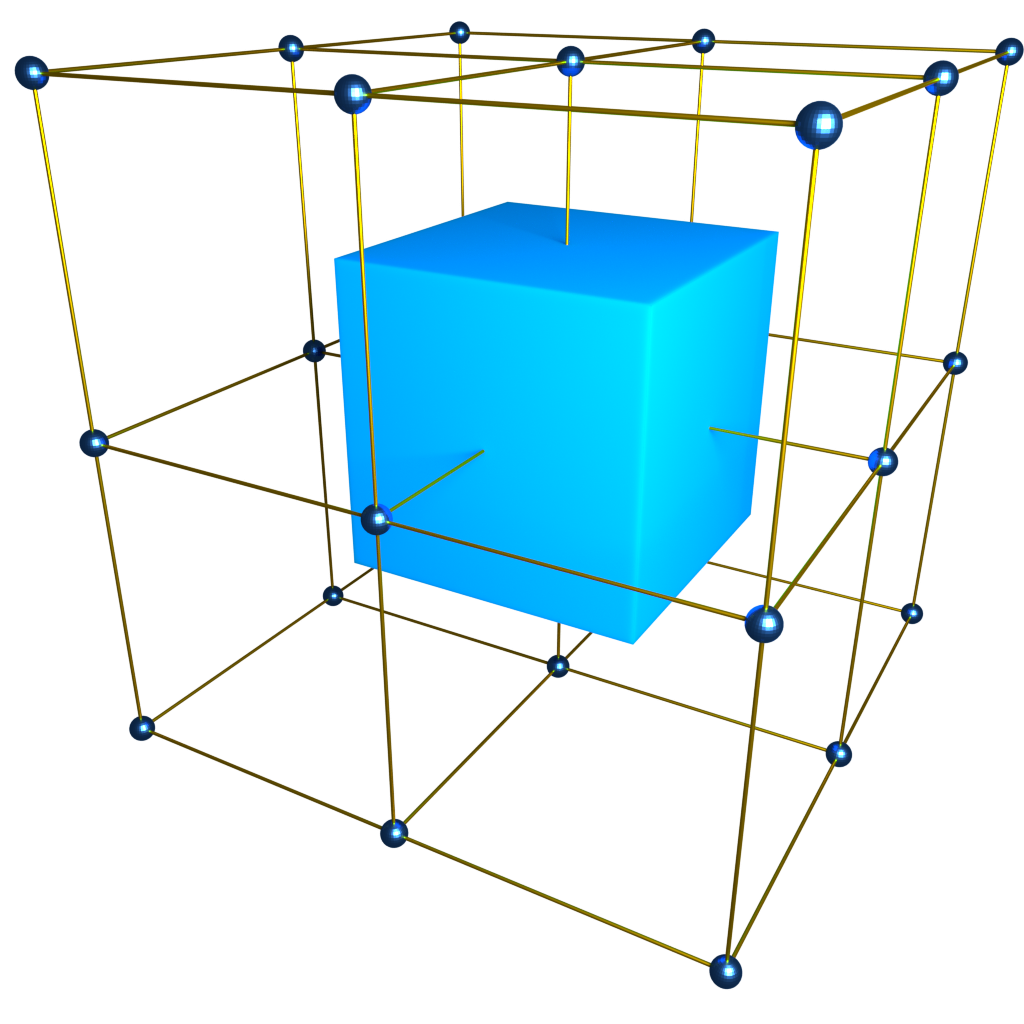}
    \includegraphics[scale=0.135]{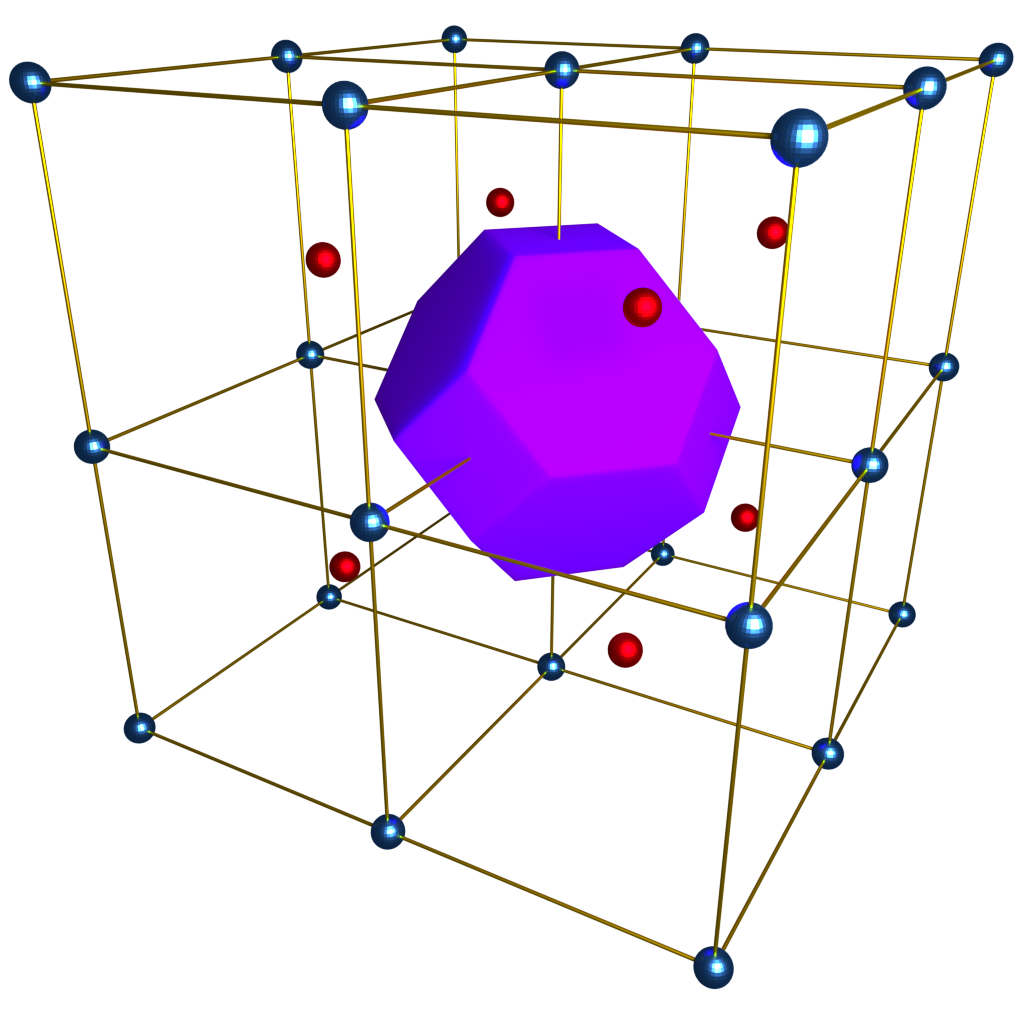}
    \includegraphics[scale=0.135]{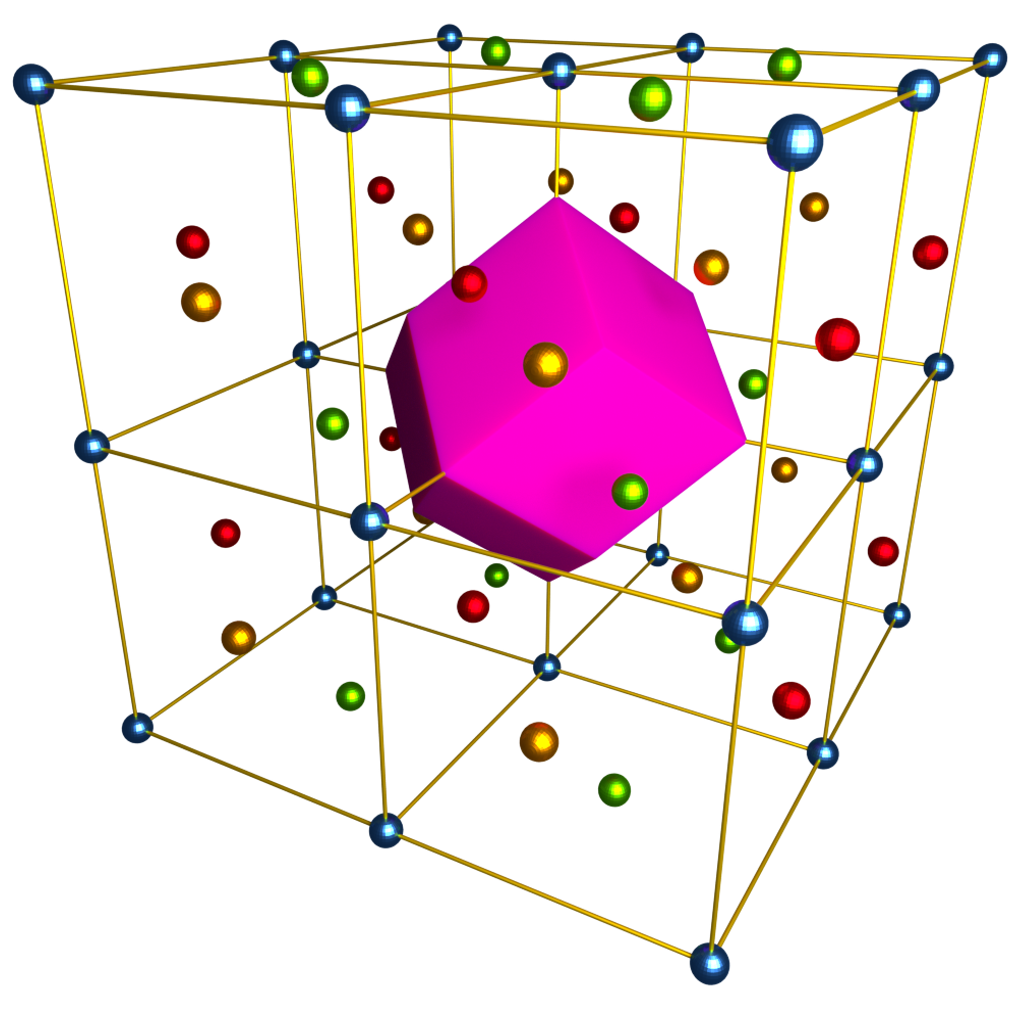} \\
    \caption{Three trivariate integer lattices. In order from left to right we have the Cartesian (CC), Body Centered Cubic (BCC) and Face Centered Cubic. The different colored lattice sites correspond to the Cartesian cosets within each sampling lattice. The polytope in the center of each grid corresponds to grid's respective Voronoi cell.}
    \label{fig:lattice_ind}
\end{figure*}

Before moving further, we need to clarify what we mean by {\em non-Cartesian computing}. We use the term ``non-Cartesian computing'' to indicate a move away from tensor-product extensions. This can take the form of non-Cartesian lattices --- take Figures~\ref{fig:lattice_ind} and~\ref{fig:2d_lattices} as examples that show the typical cubic Cartesian (CC) and square lattices in three and two dimensions, as well as more efficient non-Cartesian variants, namely the Body Centered Cubic (BCC) lattice, and the Face Centered Cubic (FCC) lattice, both of which have received a good amount of attention in the field of scientific visualization.
We may also use the term ``non-Cartesian computing'' to denote the move away from tensor-product interpolants to non-separable interpolants. In the Cartesian world, one typically uses tensor product B-splines as basis functions; however, non tensor-product splines exist, and have been investigated in the literature~\cite{retailor}. One example is the ZP-element which requires fewer memory accesses per reconstruction than the cubic tensor-product B-spline but has similar approximation properties and provides smoother, more isotropic reconstructions~\cite{entezari2006extensions}. When comparing different sampling lattices,
one may argue that a true comparison for these lattices would be their Voronoi splines~\cite{mirzargar2010voronoi}, which, until this work, have yet to have a dedicated GPU implementation.

What do we gain by adding additional complexity to our interpolation and data processing schemes? The seemingly benign difference in lattice and basis function changes the properties of the resulting approximation scheme. In fact, in 2D, an optimal hexagonal sampling scheme (Fig.~\ref{fig:2d_lattices}) yields $14\%$ better approximations when combined with an appropriate interpolant, whereas on the BCC lattice (Fig.~\ref{fig:lattice_ind})--- the optimal sampling lattice in 3D --- the improvement is around $30\%$~\cite{entezari2006towards}. Despite these advantages, the benefits of non-Cartesian computing have remained elusive. This is largely due to the complex algebro-geometric interplay between the lattices and the basis function they are paired with. These basis functions are typically composed of box splines which have intricate piece-wise polynomial structures which are challenging to evaluate in an efficient manner. Most treatments have dealt with evaluation and implementation issues in a somewhat adhoc manner.

Our goal in this work is to make the advantage of non-Cartesian computing more tangible --- explicitly, we want to make the use of alternative interpolants/lattices more practical and usable. Our main contribution is a holistic analysis of approximation spaces with a careful focus on fast implementation. We outline a  framework for translating interpolation bases into fast interpolation schemes for piece-wise polynomial interpolants. We then show how to use this pipeline to create unified \emph{fast} implementations of many of the interpolants in the non-Cartesian volumetric visualization literature. It is important to mention that we obtain a form of evaluation that is agnostic to platform. We could generate CPU code, GPU code, or Verilog from our intermediate representation. Our GPU implementation takes advantage of the in-built tensor product linear filtering hardware (i.e. linear texture fetches) on contemporary GPUs to reduce the amount of memory accesses needed for reconstruction, however this does not always lead to optimal results. 

We provide the theoretical analysis in this paper, and in the subsequent work we provide extensive implementation details. In our supplementary material we share the worksheet we use to automate the analysis as well as the code generation module, and pre-generated CUDA PTX code. To summarise, in this part, our contributions are as follows:
\begin{itemize}
    \item We provide a unified framework of analyses for combinations of splines and lattices from the context of scientific visualization.
    \item  We show how to use trilinear interpolation to reduce memory fetches and increase performance of our implementations.
    \item We provide implementations for many interpolants in the literature that have not yet had robust GPU implementations.
    \item We show the relative performance between GPU implementations, and show generality by moving towards a 4 dimensional example.
\end{itemize}

The remainder of this paper is organized as follows. In Section~\ref{sec:related_work}, we review the literature in our native context, scientific visualisation, but we will also pay note to some of the work done in one dimension and in image processing. In Section~\ref{sec:background}, we provide the background necessary for our framework. In Section~\ref{sec:method}, we attack the problem at its root, the convolution sum. We manipulate the convolution sum until it emits a form suitable for fast evaluation --- this can be thought of as unrolling the convolution sum loop. In Section~\ref{sec:results} we generate code for various interpolants (box splines and Voronoi splines) some of which have not yet appeared in the literature. We then evaluate their performance on the CPU and GPU with respect to the application of volume rendering.

\section{Related Work} \label{sec:related_work}
Interpolation is a fundamental operation in scientific visualization. Many visualization tasks (volumetric rendering, iso-surface extraction, flow visualization etc.) rely on the interpolation of a continuous function from a discrete set of values. In one dimension, the family of B-splines are the prototypical interpolant --- the linear B-spline is the least computationally expensive in this family, and has received the most attention in practice~\cite{unser1999splines}. However, it is well understood what B-splines are members of the maximal order minimal support (MOMS) family, which are the optimal interpolants for a given support~\cite{blu2001moms}. This is not the only option though, there are other constraints one may wish to design around. For example: arithmetical complexity; total number of memory accesses (i.e. support) needed for a single point reconstruction; and smoothness are all parameters that can be tweaked. If one requires infinite differentiability, then CINAPACT splines are an appropriate interpolant candidate~\cite{cinapact}. 

The design space for uni-variate interpolants is well understood. Appropriate choice of interpolant is not so clear in higher dimensions. For the domain of scientific visualization, the move to three dimensions allows practitioners the freedom to consider non-seperable splines. In this case, it is clear that the tensor product B-splines are no longer MOMS splines; the generalized ZP-element shows this, since it has lower support than (but the same approximation order as) the cubic B-spline~\cite{entezari2006extensions}.

There are multiple works that investigate non-separable basis functions on non-Cartesian lattices; box splines and weighted linear combinations of box splines often appear as popular candidates~\cite{deboorbox, domonkos2010dc, csebfalvi2013cosine}. However, it is fairly well known that evaluating a box spline numerically is quite difficult; the recursive form for box splines is unstable if naively implemented~\cite{boxeval}. One can rectify this by ensuring that any decision made while evaluating a point on the spline's separating hyperplanes is consistent~\cite{boxnumer}. Even then, recursive box spline evaluation is unsuitable for use on the GPU --- the conditional nature and large branching behaviour of the recursive form will lead to {\em branch divergence} and stall execution units on the GPU, leaving its resources underutilized. It is more convenient to work with either the Bernstein B\'ezier (BB) form, or the explicit piece-wise polynomial (PP) form of a box spline which has been characterized in closed form~\cite{horacsek2016}. Mathematically these two forms are equivalent, the BB form is simply one specific factorization of the PP form that lends itself to evaluation with De Castlejau's algorithm. Generally, a polynomial can be factorized in many different ways. 

The BB form has been used in fast GPU evaluation schemes for box splines on non-Cartesian lattices. In the work of Kim \textit{et al.}~\cite{kim2017}, the symmetry of a box spline is used to create look-up tables of the BB polynomial coefficients. Since these coefficients are rational, they are stored as pairs of integers and any division occurs at run time. This is particularly convenient on the GPU, as it allows one to write one function for De Castlejau's algorithm, and each separate region of the spline to be evaluated is a set of integers that can be looked up based on an analysis of the regions of the spline's mesh at runtime. However, this is somewhat wasteful since most regions within a spline's mesh are related by a rotation and/or a shift (i.e. an affine change of variables). As we will see in Section~\ref{sec:method}, this change of variables allows us to re-write the polynomial within the region of evaluation as a transformation followed by a polynomial evaluation of some chosen reference polynomial (provided the spline has an appropriate structure). This reference polynomial can be evaluated in BB form, however, we choose to use a Horner factorization of the PP form --- this allows us to reduce the amount of operations needed to calculate a given polynomial~\cite{horner1}. This is a generalization of the approach used in the work of Finkbeiner \textit{et al.}~\cite{finkbeiner2010efficient}.

On the BCC lattice multiple box splines have been investigated for volumetric visualization. The linear and quintic box splines were among the first used by Entezari \textit{et al.}~\cite{pracbox,entezari2009quasi} for volumetric data visualization. These splines are particularly interesting as they mimic the geometry of the BCC lattice and they attain the same order of approximation as the trilinear and tricubic splines on the CC lattice, but with smaller support. Then there is the quartic box spline of Kim~\textit{et al.} which has even smaller support than the quintic box spline, but the same order of approximation~\cite{kim2013quartic}. Additionally, in that work, Kim~\textit{et al.} proposed a 12-direction box spline with tensor product structure; this box spline has a large support size, but is reasonably fast compared to the other proposed box splines due to its tensor product flavour --- in our implementation, we call the 12-direction box spline the {\em tricubic} box spline on the BCC lattice. 

The FCC lattice has not received as much attention from researchers as the BCC lattice; we are only aware of few works that investigate box splines on the FCC lattice. In particular, Kim~\textit{et al.} have investigated a six direction box spline that respects the geometry of the FCC lattice~\cite{fccbox, Kim201390}. However, it is also true that the 12-direction box-spline~\cite{kim2013quartic} is usable on the FCC lattice. The proposed splines of Csebfalvi~\textit{et al.} may also be generalized to the FCC lattice~\cite{csebfalvi2013cosine, domonkos2010dc}.

We are only aware few works that attempt the use of a non tensor-product box spline on the CC lattice for visualization. The work of Entezari~\textit{et al.} generalizes the Zwart-Powell (ZP) element of two dimensions to three dimensions. This also maintains the same order of approximation as the cubic tensor product spline, but with smaller support~\cite{entezari2006extensions}. Even though this spline tends toward higher fidelity approximations, we are not currently aware of any GPU implementation of this spline, and we provide one in the supplementary material. 

There is also the work of Csebfalvi~\textit{et al.}, where a shifted and re-weighted box spline is designed so as to map easily to linear interpolation among the cosets of the BCC lattice~\cite{csebfalvi2013cosine}. This method has the advantage of being extremely easy to implement on the GPU, has respectable reconstruction quality and runs at decent speeds compared to trilinear interpolation. There has also been some work investigating the use of splines designed for one lattice on another~\cite{retailor}. The use of direction vectors that do not correspond to principle lattice directions helps distribute frequency content more evenly in the Fourier domain.

Another intuitive idea is to look at the Voronoi cell of a lattice and convolve that with itself to obtain an interpolant, Figure~\ref{fig:lattice_ind} shows the CC, BCC and FCC lattices with their Voronoi cells. This produces a valid approximation scheme, but is quite expensive to compute~\cite{mirzargar2010voronoi,van2004hex,mirzargar2011quasi}. Until this work, there was no GPU implementation available for these splines. These fit within our pipeline, and we provide an implementation for the BCC and FCC lattices --- Voronoi splines on the CC lattice correspond to the tensor product B-spline. 

Finally, while not strictly related to spline evaluation, it is important to discuss {\em quasi-interpolation}, since an approximation space may not harness the full approximation power of a given basis function without proper pre-filtering. Most of the basis functions discussed so far are not interpolating. If a basis is stable (\textit{i.e} it forms a {\em Riesz basis}) then it is possible to process the input data so that the resulting reconstruction interpolates data~\cite{usmanThesis}. However, this does not always yield the best reconstruction. Moreover, some bases cannot be made interpolating, so what is to be done in those cases? Quasi-interpolants ensure that a reconstruction harnesses the full approximation order of a space --- in general it is a good idea to prefilter data with a quasi-interpolant if a basis is not interpolating. This is related to an error kernel analysis, where the data convolved with a filter is guaranteed to decay with the order of the basis~\cite{blu1999quantitative, usmanThesis}. While this is an important ingredient to a good reconstruction scheme, in this work we are not concerned with the approximation properties of a basis function, only on the speed at which it can be evaluated. 

\section{Background} \label{sec:background}
When we work with data that are uniformly sampled in one dimension there is only one possible sampling scheme --- take equally spaced samples along the indepenent axis. The straightforward extension of this to multiple dimensions is to sample evenly in all cardinal directions, i.e. a Cartesian sampling. However, the situation is generally more complex in higher dimensions.

\begin{figure}
    \centering
    \includegraphics[]{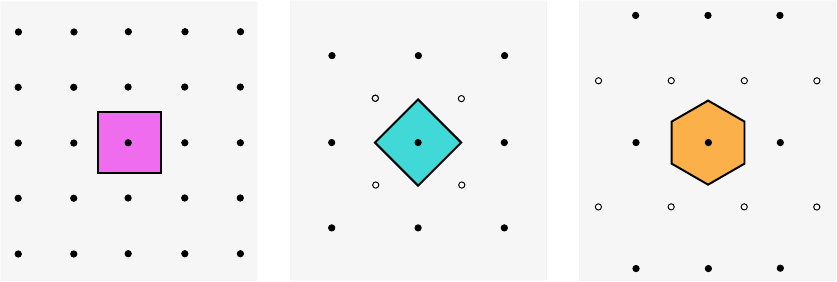}
    \caption{The square Cartesian (left) quincunx (middle) and hexagonal (right) lattices respectively. Notice the coset structure of the quincunx lattice is two interleaved Cartesian lattices --- denoted by the difference in lattice site coloring. The same applies for the hexagonal lattice, although slightly distorted.}
    \label{fig:2d_lattices}
\end{figure}

To move away from Cartesian lattices and account for this complexity, a function can be sampled according to an invertible matrix. This matrix --- known as the generating matrix --- represents a lattice structure. That is, if we have a real valued function $f(\mathbf{x})$ in some dimension $s$, we define the sampled version of that function with respect to a given lattice as $f_{\mathbf{n}} := f(hL\mathbf{n})$. Here, $h$ is a real valued scale parameter that controls the granularity or coarseness of the sampling pattern (i.e. the sampling rate), $\mathbf{n} \in \mathbb{Z}^s$ and $L$ is the generating matrix of the lattice. In any dimension, the $s \times s$ identity matrix generates a Cartesian sampling. In two dimensions the hexagonal and quincunx lattices are generated by the matrices
\begin{equation}
    L_{HEX}:=\begin{bmatrix}
        1 & \frac{1}{2} \\
        0 & \frac{\sqrt3}{2}
    \end{bmatrix}
    \;\; \text{and} \;\;
    L_{QC}:=\begin{bmatrix}
            1 & -1 \\
            1 & 1
    \end{bmatrix}
\end{equation}
respectively (Figure~\ref{fig:2d_lattices}). In three dimensions, the BCC and FCC lattices are generated by
\begin{equation}
    L_{BCC}:=\begin{bmatrix}
        -1 & \phantom{-}1 & \phantom{-}1 \\
        \phantom{-}1 & -1 & \phantom{-}1 \\
        \phantom{-}1 &  \phantom{-}1 & -1
    \end{bmatrix}
    \;\; \text{and} \;\;
    L_{FCC}:=\begin{bmatrix}
        0 & 1 & 1 \\
        1 & 0 & 1 \\
        1 & 1 & 0
    \end{bmatrix}
\end{equation}
respectively (Figure~\ref{fig:lattice_ind}). It is worth noting that a lattice may be generated by any number of different generating matrices --- one may construct an equivalent lattice by carefully choosing vectors in $L\mathbb{Z}^s$. In general, finding a minimal basis for a given generating matrix is an NP-hard problem~\cite{conway2013sphere}, as such it is difficult to find a relationship between matrices that generate the same lattice.

While lattice theory is quite deep and general, we only adhere to the case where $L\in\mathbb{Z}^{s\times s}$ and $L$ has full rank. This is because, in practice, we use lattices as array-like data structures and it is not possible to index via real valued numbers. As a consequence, it is not possible for us to directly represent functions on the hexagonal lattice within our framework. However, it is possible to stretch an integer lattice into another non-integer lattice; for example, in Figure~\ref{fig:2d_lattices}, the hexagonal lattice can be seen as a distorted quincunx lattice. 

Another fundamental object related to a lattice is its {\em Voronoi} region, which is the set of points closest to the lattice site situated at the origin (both Figure~\ref{fig:lattice_ind} and Figure~\ref{fig:2d_lattices} show each lattice's Voronoi region). The Voronoi region of a lattice embodies all the properties of a lattice --- from it one may derive the shortest vector of a lattice (i.e. the shortest vector problem which is NP-complete) as well as set of shortest spanning vectors, the symmetry group of the lattice, packing efficiency etc. ~\cite{conway2013sphere,viterbo1996computing}. Unsurprisingly, computing the Voronoi region of a lattice is NP-complete as well, but for reasonably specified bases in lower dimensions, the problem is tractable via the Diamond Cutting algorithm~\cite{viterbo1996computing}.

To reconstruct a function on a given lattice, we choose a basis function $\varphi(\mathbf{x})$, shift it to each lattice site, and modulate it by the coefficient stored at that lattice site. The space of all functions spanned by the lattice shifts of $\varphi$ is defined as the set
\begin{equation}\label{eq:conv_sum}
    \mathbb{V}\left(\varphi,L,h\right) := \left\{ \sum_{\mathbf{n}\in L\mathbb{Z}^s}c_{\mathbf{n}}\varphi\left(\frac{\mathbf{x}}{h}-\mathbf{n}\right) : \mathbf{c} \in l_2 \right\},
\end{equation}
where $\mathbf{c}$ is a real-valued coefficient sequence associated with the lattice. Obtaining a good approximation $\tilde{f}(\mathbf{x}) \in \mathbb{V}(\varphi,L,h)$ to $f(\mathbf{x})$ equates to finding an appropriate set $\{c_\mathbf{n}\}$; this is highly dependent on the choice of $\varphi(\mathbf{x})$ and $L$. If we are so lucky that $\varphi(\mathbf{x})$ is interpolating on $L$ (i.e. $\varphi(\mathbf{0}) = 1$ and $\varphi(L\mathbf{n}) = 0$ when $\mathbf{n} \in \mathbb{Z}^s/\{\mathbf{0}\}$), then we may choose $c_{\mathbf{n}} = f(hL\mathbf{n})$. However, in general, some preprocessing must be performed in order to achieve optimal error decay between the original function and its approximation. This is beyond the scope of this article, but the interested reader may refer to some of the related work~\cite{unser2000sampling, blu1999quantitative, usmanThesis}. Since any asymptotic analysis is outside of the scope of this paper, without loss of generalization we set $h=1$. 

As for specific choices of $\varphi$, box splines have been investigated extensively in scientific visualization. Box splines have many equivalent definitions, but perhaps the most intuitive is the convolutional definition where one successively convolves an indicator function along the $\boldsymbol\xi_i$ direction vectors~\cite{deboorbox}. We collect $n$ of these vectors in an $s \times n$ matrix
\begin{equation}
    \Xi := \begin{bmatrix}
        \boldsymbol\xi_1 & \boldsymbol\xi_2 & \boldsymbol\cdots & \boldsymbol\xi_n
    \end{bmatrix}.
\end{equation}
In our case, we are interested only in matrices where $n \ge s$ and the dimension of the range of $\Xi$ is equal to $s$; we require these for a valid approximation scheme~\cite{deboorbox}. With this, a box spline can be defined recursively as
\begin{equation}
    M_\Xi(\mathbf{x}) := \int_0^1 M_{\Xi/\{\boldsymbol\xi\}}(t\boldsymbol\xi - \mathbf{x}) dt,
    \quad \text{for} \; n > s,
\end{equation}
where ${\Xi/\{\boldsymbol\xi\}}$ is interpreted as removing a direction vector $\boldsymbol{\xi}$ from the matrix $\Xi$. When $n=s$, we define $M_\Xi(\mathbf{x})$ as
\begin{equation}
    M_\Xi(\mathbf{x}) := \begin{cases} 
      \frac{1}{\det|\Xi|} & \text{if } \Xi^{-1}\mathbf{x} \in [0,1)^s  \\
      0 & \text{otherwise} ,
   \end{cases}
\end{equation}
which provides a base case for the recursion. 

The box spline $M_\Xi(\mathbf{x})$ is compactly supported within a polytope defined by the Minkowski sum of the direction vectors in $\Xi$. We denote the support as $\text{supp} \, M_\Xi$; it is given by
\begin{equation}
    \text{supp}\,M_\Xi = \left\{\sum_{k=1}^n a_k \boldsymbol{\xi}_k \;|\; \forall(k)\; a_k \in (0,1) \right\}.
\end{equation}
$M_\Xi$ is piecewise polynomial within its support. The polynomial pieces are delineated by all $(s-1)$-dimensional hyperplanes spanned by the direction vectors in $\Xi$. Denoting $\mathbb{H}$ as the set of all such hyperplanes, we can define a fine mesh that is formed by the lattice shifts of $M_\Xi$:
\begin{equation}
    \Gamma(\Xi, L) := \bigcup_{H\in\mathbb{H}} H + L\mathbb{Z}^s.
\end{equation}
Each region of the mesh is a convex polytope that contains a polynomial, typically these are different polynomials, however, it is possible that some regions may share the same polynomial. In general, it is possible to derive the explicit piece-wise polynomial form for each of these regions~\cite{horacsek2016}--- we will use this explicit form in Section~\ref{sec:method} to determine a simpler representation of the convolution sum in Equation~\ref{eq:conv_sum}. 

Our framework also accounts for Voronoi splines, which can be intuitively defined as successive convolutions of a lattice's indicator region with itself. More precisely, Voronoi splines are defined as
\begin{eqnarray}
    V_0(\mathbf{x}) &:=& \chi_L(\mathbf{x}) \\
    V_n(\mathbf{x}) &:=& (V_0 \ast V_{n-1})(\mathbf{x})
\end{eqnarray}
where $\ast$ denotes convolution, and $\chi_L(\mathbf{x})  = 1/|\det{L}|$ if $\mathbf{x}$ is in the Voronoi region of the lattice $L$, and $0$ otherwise. Voronoi splines are also piece-wise polynomial, and their piece-wise polynomial form may be obtained by first redefining $V_0(\mathbf{x})$ as the sum of a collection of constant valued box splines; higher order $V_n(\mathbf{x})$ may be obtained by the convolution definition~\cite{mirzargar2010voronoi}. Unrolling the recursion, we obtain $V_n(\mathbf{x}) = V_0(\mathbf{x})^{\ast n}$ where $\ast n$ denotes successive convolutions. If $V_0(\mathbf{x})$ is the sum of box-splines, then one may use the multinomial theorem to expand $V_0(\mathbf{x})^{\ast n}$ into the sum of many box-spline convolutions; convolving a box-spline with another box-spline simply produces a (possibly shifted) box-spline. Thus, by summing all these box-splines, we arrive at the final piece-wise polynomial form. However, these calculations are not trivial. For the BCC lattice, $V_0(\mathbf{x})$ is the sum of 16 box-splines, $V_1(\mathbf{x})$ (analogous to the tri-linear B-spline on the CC lattice) is the sum of 136 box-splines, $V_2(\mathbf{x})$ is the sum of 816 and $V_3(\mathbf{x})$, the sum of 3876~\cite{mirzargar2010voronoi}. While these numbers are not intractably large, keep in mind that one must compute a box-spline's polynomial representation at every iteration, which can take a non-trivial amount of time; on the order of a few dozen minutes, for the splines that sum to $V_3(\mathbf{x})$. Spread across 4 CPU cores, this took us about a month to compute.

\section{Methodology} \label{sec:method}
\begin{figure}[t!]
    \centering
    \includegraphics[]{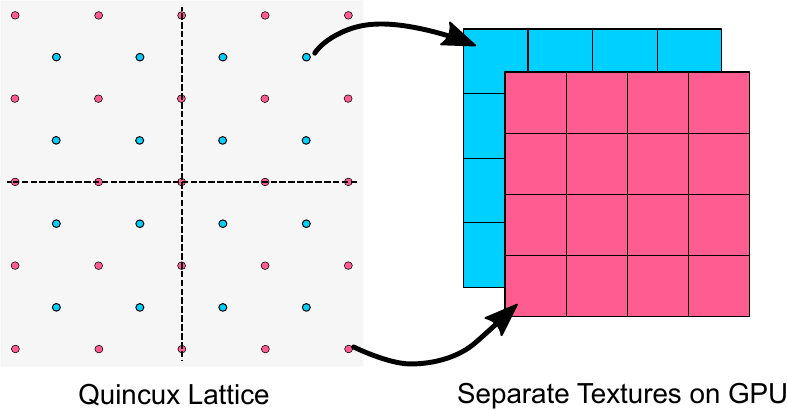}
    \caption{Lattices are broken into their Cartesian coset structure, then stored in separate $s$-dimensional arrays. On the GPU, this translates to a collection of textures. For volumetric data, we decompose the lattice into separate volumetric textures. 
    } \label{fig:texture_format}
\end{figure}

The convolution sum in Equation~\ref{eq:conv_sum} is the focus of this section. There are two important parts to evaluating a function of this form; the first is the representation of the coefficient set $c_{\mathbf{n}}$ (i.e. the memory layout), the second is the evaluation of the basis function $\varphi$. Concerning $c_{\mathbf{n}}$, we choose a specific representation in which each $c_{\mathbf{n}}$ corresponds to a fetch from an $s$-dimensional array. Thus, for any lattice, we choose to treat it as a union of shifted Cartesian lattices (also known as a Cartesian coset decomposition). This allows us to use texture memory on GPUs to store each coset, as shown in Figure~\ref{fig:texture_format}. Provided we combine all reads from a given coset, this also allows us to take advantage of data locality (i.e. cache). It also allows us to use trilinear fetches as building blocks for our interpolation schemes --- that is, we attempt to write our interpolation schemes as weighted sums of trilinear texture fetches on the different cosets of the lattice. This does not necessarily reduce computation per evaluation, but rather reduces the number of texture fetches needed to evaluate the convolution sum, and therefore reduces bandwidth. 

\begin{figure}[t!]
    \includegraphics[width=0.25\textwidth]{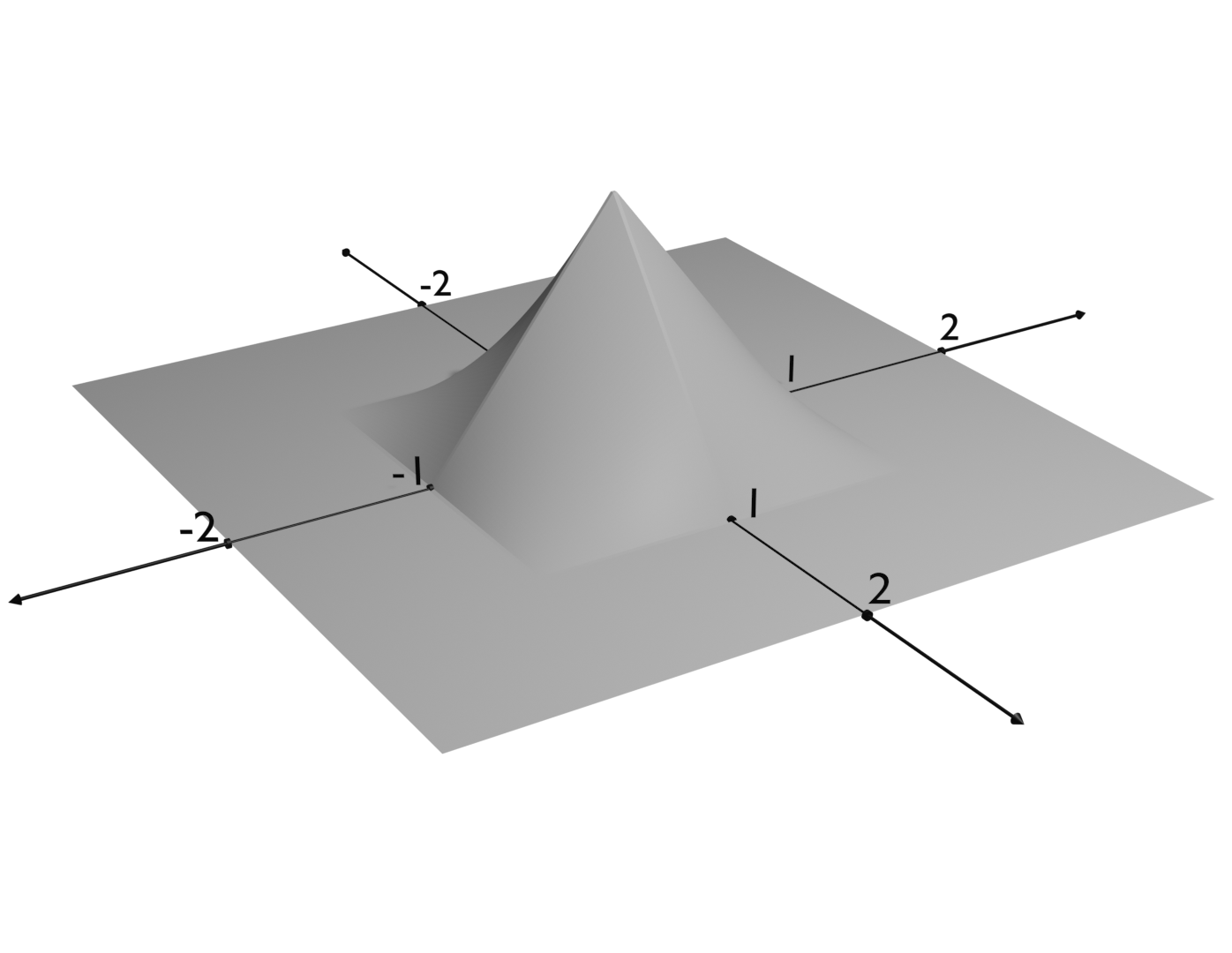} \
    \includegraphics[width=0.25\textwidth]{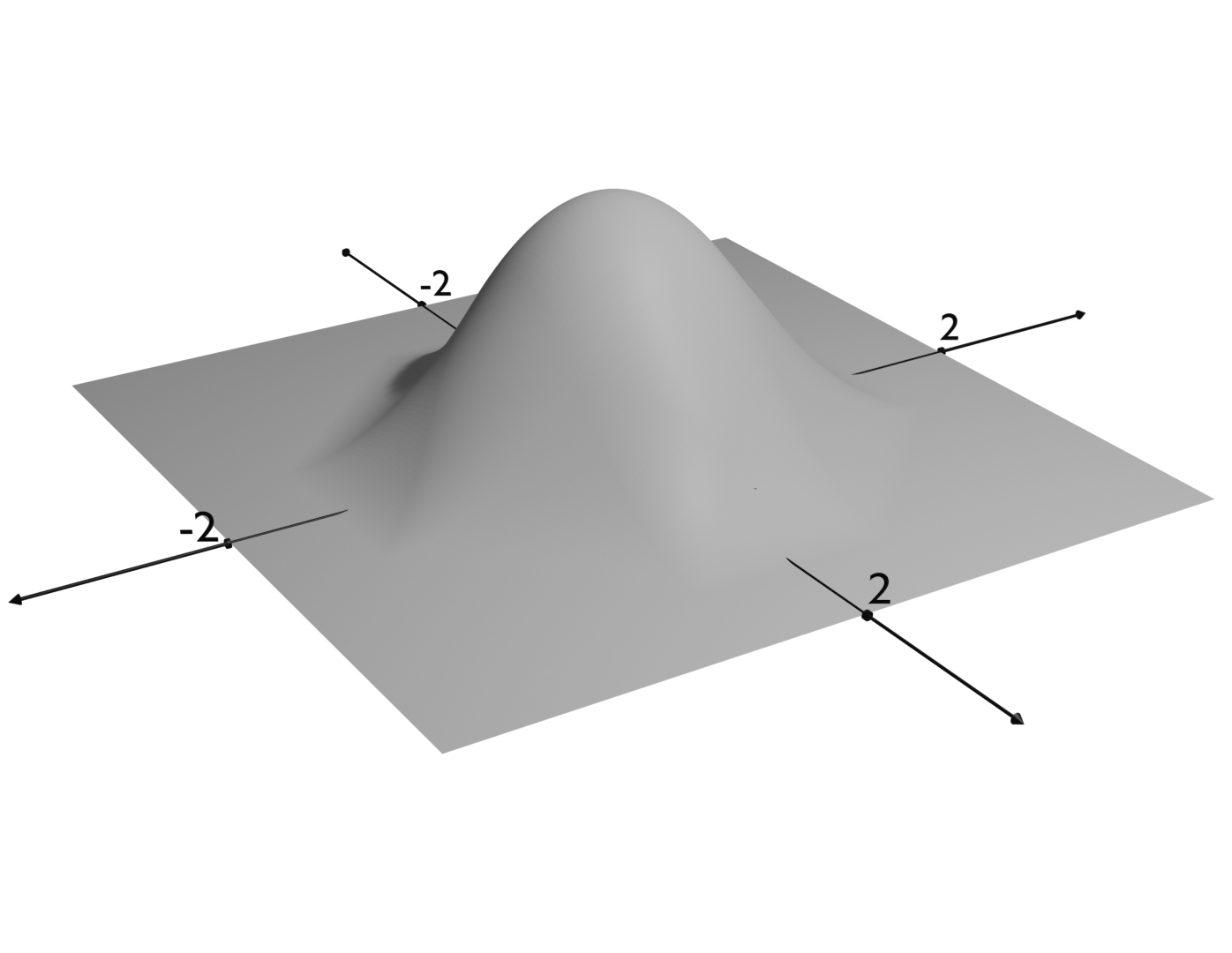} \
    \includegraphics[width=0.25\textwidth]{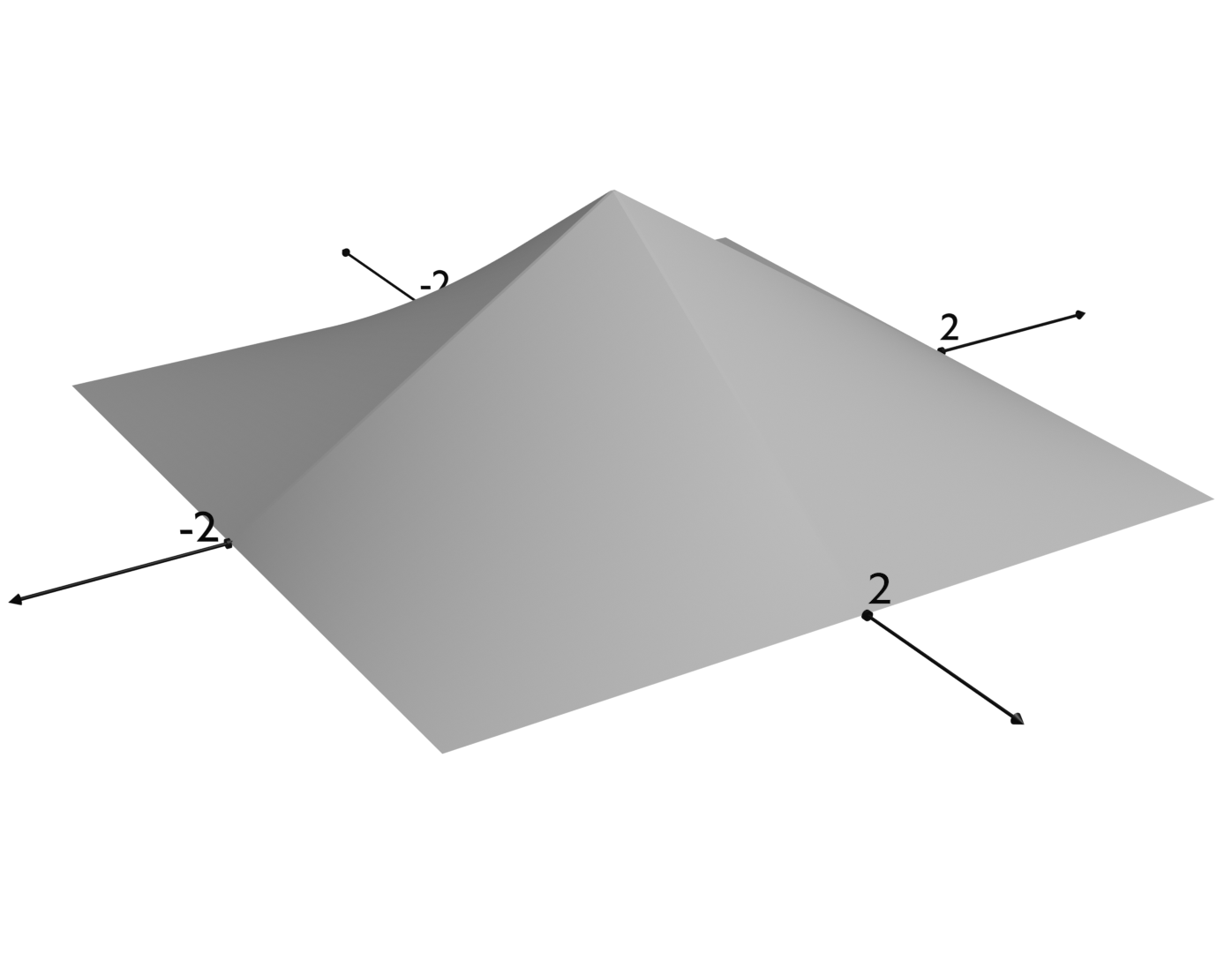}
    \caption{Splines used as running examples; from left to right are box splines generated by $\Xi_{TP2}, \Xi_{ZP}$ and $\Xi_{QC}$ respectively. 
    } \label{fig:2dsplines}
\end{figure}

With the memory layout solidified, we turn to the basis function --- we apply a series of manipulations to ``unroll'' the convolution sum. We build some insight into how to transform $\varphi$ into a more convenient form for evaluation with some running examples. We consider the linear tensor product spline and the Zwart-Powell (ZP) element on the two-dimensional Cartesian lattice. These have direction matrices
$$
     \Xi_{\text{TP2}} = \begin{bmatrix} 1  &0 &-1& 0\\ 0 & 1 & 0 & -1 \end{bmatrix},
     \Xi_{\text{ZP}} = \begin{bmatrix} 1  &0 & 1& 1\\ 1 & 1 & 1 & -1 \end{bmatrix}
$$
respectively. On the quincunx lattice, we use a tensor product style spline defined by 
$$
    \Xi_{\text{QC}} = \begin{bmatrix} 2  &0 &-2& 0\\ 0 & 2 & 0 & -2 \end{bmatrix}.
$$
These box splines are shown in Figure~\ref{fig:2dsplines}. While this is nowhere near an exhaustive list of interpolants, the examples we provide are simple enough to understand in 2D and show the intricacies of the procedure.

\subsection{Manipulating the Convolution Sum}
We start by reiterating the convolution sum in Equation~(\ref{eq:conv_sum}):
\begin{equation}
    f(\mathbf{x}) := \sum_{\mathbf{n} \in L\mathbb{Z}^s} c_{\mathbf{n}}\varphi(\mathbf{x}-\mathbf{n}).
\end{equation}
At a high level, our methodology consists of incrementally applying a series of simple algebraic manipulations to this sum so that it can be more effectively mapped to an implementation; be it CPU or GPU. Along the way, we pause and discuss the effects of certain choices and properties of $\varphi$ on an evaluation scheme. 

The first manipulation we apply is one that has been used to derive fast interpolants on the BCC and FCC lattices~\cite{csebfalvi2013cosine,domonkos2010dc}. In particular, these interpolation schemes take advantage of the fact that an integer lattice can be decomposed into a sum of sub-lattices. We build upon this observation by noting that we may decompose the convolution sum up over \emph{any} coset structure of the lattice. This decomposition is independent of our data representation. For example, one may take a cubic Cartesian lattice and decompose it as two shifted BCC lattices. Within our framework, on the GPU each of these two BCC lattices will in turn have two 3D textures associated with it.

 Formally, if $G$ is an integer subsampling matrix that yields a sub-lattice of $L$, then there exist $M := |\det G|$ integer vectors $\mathbf{l}_0, \mathbf{l}_1,  \cdots, \mathbf{l}_{M-1}$ such that $L\mathbb{Z}^s = \cup_{k=0}^{M-1} GL\mathbb{Z}^s+\mathbf{l}_k$. Note that the lattice sites $GL\mathbb{Z}^s + \mathbf{l}_k$ constitute the coset corresponding to the shift vector $\mathbf{l}_k$. Without loss of generality, we take $\mathbf{l}_0$ to be zero vector of $\mathbb{Z}^s$. We can now write the convolution sum as
\begin{equation}
    \sum_{\mathbf{n} \in L\mathbb{Z}^s} c_{\mathbf{n}}  \varphi(\mathbf{x}-\mathbf{n}) = 
    \sum_{k=0}^{M-1}\sum_{\mathbf{m} \in GL\mathbb{Z}^s+\mathbf{l}_k} c_{\mathbf{m}}\varphi(\mathbf{x}-\mathbf{m}),
\end{equation}
which decomposes the sum over the coset structure of the lattice induced by $G$. Figure~\ref{fig:2d_lattices} shows the Cartesian coset structure of the quincunx lattice. In three dimensions, the BCC lattice emits similar behaviour to this; it consists of two interleaved Cartesian grids where one of the Cartesian grids is shifted by the vector $(1,1,1)$. The FCC lattice is similarly decomposed into Cartesian cosets, as shown in Figure~\ref{fig:lattice_ind}.

The matrix $G$ is a parameter choice that must be made depending on what is most appropriate for the device on which we are implementing an interpolation scheme. Currently, for evaluation on the GPU, a reasonable choice is a $G$ such that $GL=D$ where $D$ is a diagonal matrix. In other words, $G$ yields a Cartesian coset structure, and we may store the lattice as a collection of Cartesian lattices, specifically 3D textures (see Figure~\ref{fig:texture_format}). However, it is not always strictly advantageous to do so; if $\varphi$ does not have partition of unity on the sub-lattice generated by $GL$, then this complicates the geometric decomposition of the spline. We will revisit this when we discuss the {\em sub-regions of evaluation} of a spline. While an integer invertible $G$ will still produce a valid evaluation scheme, in practice $G$ must be chosen so that it both produces a simple sub-lattice $GL$ and respects the geometry of $\varphi$.

Next, we note that in all practical instances we have a compact basis function $\varphi$ --- either we choose $\varphi$ to be compact, or the finite nature of our data imply a bound on the support of $\varphi$. Thus we change our perspective and rewrite the convolution sum over the support of $\varphi$ shifted to the evaluation point $\mathbf{x}$. From this perspective, any lattice site that falls within the support of this function will contribute to the reconstruction. To simplify notation, we define the set 
\begin{equation}
    C_k(\mathbf{x}) := \text{supp} \ \varphi(\cdot -\mathbf{x}) \cap (GL\mathbb{Z}^s + \mathbf{l}_k).
\end{equation}
As such, $C_k(\mathbf{x})$ is the set of lattice sites on coset $k$ that contribute to the reconstruction at point $\mathbf{x}$. Therefore, we may write 
\begin{equation} \label{eq:coset_form}
    \sum_{{\mathbf{n}} \in L\mathbb{Z}^s} c_{\mathbf{n}}\varphi({\mathbf{x}}-{\mathbf{n}}) = 
    \sum_{k=0}^{M-1}\sum_{{\mathbf{m}} \in C_k({\mathbf{x}})} c_{\mathbf{m}}\varphi({\mathbf{x}}-{\mathbf{m}}).
\end{equation}
So far, we have not strictly required $\varphi$ to be compact, but now we impose the following constraints on $\varphi$: it has partition of unity on $L$, and it is piece-wise polynomial with compact polyhedral support. Partition of unity --- the ability for a reconstruction space to reproduce a constant function --- is a basic requirement for a valid approximation scheme, along with polynomial decay in the Fourier domain~\cite{strang2011fourier}. Compactness is generally not restrictive, almost all $\varphi$ used in practice are compact. The polynomial restriction is slightly restrictive, as it does not allow us to consider the class of exponential box splines~\cite{ron1988exponential}, the CINAPACT splines~\cite{cinapact}, or the cosine weighted spline~\cite{csebfalvi2013cosine} (which is an exponential box spline). 

From Equation~(\ref{eq:coset_form}), it is clear how the support of $\varphi$ affects the inner summation of this equation --- if there are more lattice sites within the support of $\varphi$, then the inner summation will run over more terms. The effect this has on performance depends on the underlying architecture of the machine and the complexity of $\varphi$. If we can commit memory reads for the $c_{\mathbf{m}}$ while beginning to compute parts of $\varphi$ that do not depend on the coefficients of the summation, we can effectively hide the computation of $\varphi$ in the latency of the memory fetches. Latency hiding of this form is common in GPU compute applications. Moreover, the out-of-order execution units on modern CPUs allow for this behaviour without explicitly coding for it. However, if memory accesses are not an issue (i.e. if we are reading from fast memory, such as cache), then the cost of reconstruction will be dominated by the complexity of $\varphi$, and it would be advantageous to choose $\varphi$ with larger support but lower evaluation time complexity. 

\begin{figure}[t!]
    \centering
    \includegraphics[]{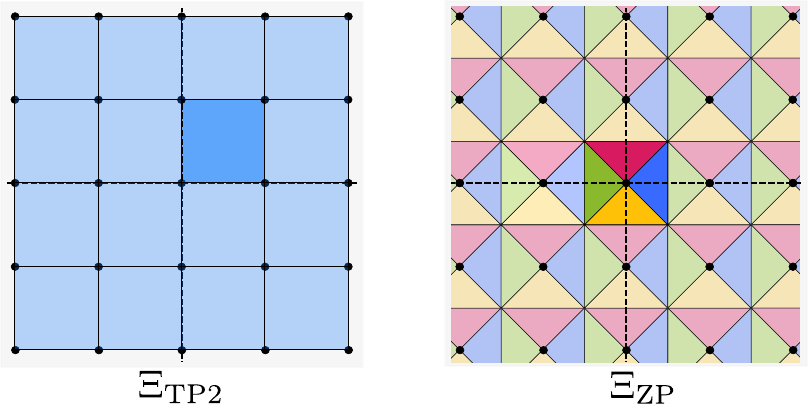}
    \caption{Sub-regions and regions of evaluation on the Cartesian lattice for the tensor product box splines defined by $\Xi_{\text{TP}n}$ and the ZP element defined by $\Xi_{\text{ZP}}$.  The equivalence classes for the sub-regions of evaluation are denoted by the hue of the region. The sub-regions are collected near the origin to create the region of evaluation.}
    \label{fig:regions_cc}
\end{figure}

We now shift our focus to the geometry of $\varphi$. Without loss of generality, we limit our discussion to $C_0(\mathbf{x})$ since all other cases are  shifted versions of this base case. Let us choose some $\mathbf{x}$ such that there exists some $\epsilon$-neighborhood around $\mathbf{x}$ for which $C_0(\mathbf{x})$ remains constant. Then there is some larger region $S$ containing $\mathbf{x}$ for which $C_0(\mathbf{x})$ does not change. That is, if we pick any $\mathbf{x,y}\in S$ then $C_0(\mathbf{x}) = C_0(\mathbf{y})$. The closure of these regions tessellate space; Figure~\ref{fig:regions_cc} shows examples of this. For box splines, these correspond to the regions within the finer mesh $\Gamma(\Xi, GL)$ of a the box spline on the lattice $GL$~\cite{deboorbox}. In this work, we refer to these as the {\em sub-regions} of the spline.

\begin{figure}[t!]
    \centering
    \includegraphics[]{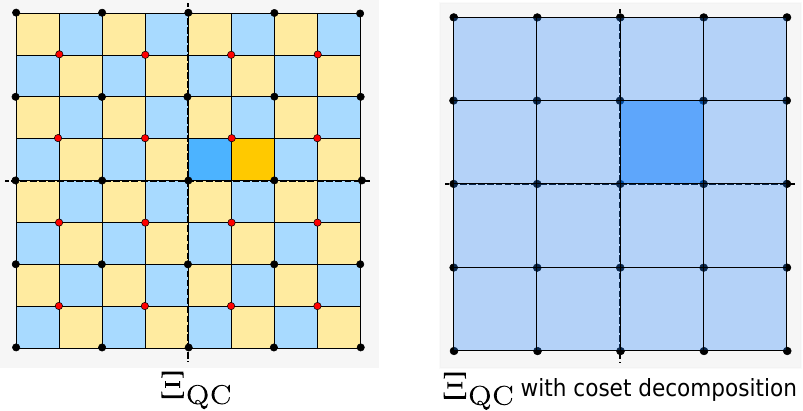}
    \caption{Sub-regions and regions of evaluation on the quincunx lattice for the tensor product box spline with no coset matrix, and coset matrix $G=L_{QC}^T$.  At the origin are the representative sub-regions of evaluation. Again, the equivalence classes for the sub-regions of evaluation are denoted by the hue of the region. In this figure, note the region of evaluation for the example on the left, shifting a point of evaluation to this region requires more logic than the case in which we use an appropriate coset decomposition, as seen on the right.}
    \label{fig:regions_qc}
\end{figure}

Note the periodic nature of the sub-regions. They naturally fit into equivalence classes under the following definition: we say that two mesh regions $P$ and $Q$ belong to the same equivalence class if for any $\mathbf{x}\in P$, there exists some lattice shift $\mathbf{k} \in GL\mathbb{Z}^s$ such that $\mathbf{x} + \mathbf{k} \in Q$. Note that there are only finitely many equivalence classes, say $N$ of them, and we denote them as $S_0, S_1, \cdots, S_{N-1}$. We will call the choice of representative mesh regions {\em sub-regions of evaluation}. We define a {\em region of evaluation} as a convex polyhedral collection of sub-regions with hyper-volume $|\det{GL}|$. Note that, since we have hyper-volume $|\det{GL}|$, a region of evaluation will tessellate space with one convex polyhedron about the lattice $GL$. Additionally, while the sub-region equivalence classes are determined by the (decomposed) lattice and $\varphi$, there are multiple possible regions of evaluation for a given $GL$ (see Figure~\ref{fig:regions_cc} and Figure~\ref{fig:regions_qc}). We choose all our regions of evaluation so that they contain or touch the origin, we call the choice of such a region $R$.


The reason we focus on a single region of evaluation $R$ is that it allows us to exploit the shift invariant nature of the approximation space. That is, say we derive a fast evaluation function $\zeta(\mathbf{x})$ with $\zeta(\mathbf{x})=\sum_{k=0}^{M-1}\sum_{{\mathbf{m}} \in C_k({\mathbf{x}})} c_{\mathbf{m}}\varphi({\mathbf{x}}-{\mathbf{m}})$ for $\mathbf{x}\in R$ (but perhaps not outside $R$). The function $\zeta $ necessarily depends on some subset of coefficients $\{c_{\mathbf{i}} : \mathbf{i} \in A\}$ for some fixed $A \subset \mathbb{Z}^s$. Since the region of evaluation tessellates space, for any $\mathbf{x} \in \mathbb{R}^s$ there exists some $\mathbf{k} \in GL\mathbb{Z}^s$ such that $\mathbf{x} - \mathbf{k} \in R$. Thus, we can derive a fast evaluation scheme for all points by shifting the evaluation to $\zeta(\mathbf{x}-\mathbf{k})$ and shifting the coefficients $\{c_{\mathbf{i}+\mathbf{k}} :\mathbf{i} \in A\}$ --- this is a consequence of the ``shift invariance'' property of shift invariant spaces. We define the function 
\begin{equation}
    \rho : \mathbb{R}^s \rightarrow GL\mathbb{Z}^s \text{ such that } \mathbf{x}\in \mathbb{R}^s \implies \mathbf{x}-\rho{(\mathbf{x})} \in R
\end{equation}
to formalize this notion, and assert that this must exist if $R$ tessellates space and has hypervolume equal to $|\det GL |$.

Thanks to this shift invariance, we may focus solely on the region of evaluation. To each representative sub-region of evaluation, we associate a function $ \psi_{j}(\mathbf{y}, \mathbf{k})$ that is defined as follows:
\begin{equation}
    \psi_{j}(\mathbf{y}, \mathbf{k}) :=  \sum_{\mathbf{n} \in C_0(S_j)} c_{\mathbf{n}+\mathbf{k}}\varphi(\mathbf{y}-\mathbf{n}) \quad  \text{for }  \quad \mathbf{y} \in S_j.
\end{equation}
For our piece-wise polynomial splines, we may explicitly construct the polynomial representation of 
$\psi_{j}(\mathbf{y}, \mathbf{k})$. We use a greedy Horner scheme to optimize the evaluation of  $\psi_{j}(\mathbf{y}, \mathbf{k})$, computation on the GPU is relatively cheap, at least compared to bandwidth access.

We combine all these sub-region functions to a single ``approximation'' function,  $\Psi(\mathbf{x}, \mathbf{l})$, with the definition
\begin{equation} \label{eq:breakdown}
    \Psi(\mathbf{x}, \mathbf{l}) := \begin{cases} 
      \psi_0(\mathbf{x}-\rho(\mathbf{x}),\rho(\mathbf{x}) +\mathbf{l}) & \text{if } \mathbf{x} - \rho(\mathbf{x}) \in  S_0  \\
      \psi_1(\mathbf{x}-\rho(\mathbf{x}),\rho(\mathbf{x}) +\mathbf{l}) & \text{if } \mathbf{x} - \rho(\mathbf{x}) \in  S_1  \\
      \ & \vdots \\
      \psi_{N-1}(\mathbf{x}-\rho(\mathbf{x}),\rho(\mathbf{x}) +\mathbf{l}) & \text{if } \mathbf{x} - \rho(\mathbf{x}) \in  S_{N-1}.  \\
   \end{cases}
\end{equation}
We can finally write the convolution sum per coset as
\begin{equation}
    \sum_{\mathbf{n} \in C_k(\mathbf{x})} c_{\mathbf{n}}\varphi(\mathbf{x} - \mathbf{n})  = \Psi(\mathbf{x}-\mathbf{l}_k, \mathbf{l}_k).
\end{equation} 
To obtain the final approximation, one only needs to sum this over all cosets.

While this is an optimized form, it still contains multiple branches --- we have different cases for each sub-region. Modern CPUs have been heavily optimized to efficiently execute branch heavy code, whereas GPUs have not. The conditional behaviour of $\Psi(\mathbf{x}, \mathbf{l})$ is a major problem for GPU implementations. However, it is possible to use the symmetry of $\varphi$ to eliminate branches. A similar idea has been used to derive fast interpolation schemes on non-Cartesian lattices~\cite{kimeval}. The approach taken in other works has been to encode each different polynomial in BB form, use the sub-regions to determine which coefficient sequence to use in deCasteljau's algorithm~\cite{kimeval}. In this work, we use a slightly more efficient idea --- we use the symmetry of the basis function to encode a change of variable relationship between subregions. This allows us to rely on a small set of polynomials, rather than having one for each sub-region. We first discuss how to determine which sub-region a point lies within without any branches. We then discuss how to eliminate branches in Equation~\ref{eq:breakdown}.

\subsection{Branch Free Sub-region Membership}
Suppose we have some $\mathbf{x} \in R$, since $R$ is a convex polyhedron, and is the closed union of finitely many convex polyhedra, we know there is a finite set of planes that divide the region of evaluation into the sub-regions of evaluation. We will say we have $Q$ of these planes, and we will call the set of planes $P=\{(\mathbf{p}_0,d_0), (\mathbf{p}_1, d_1), \cdots, (\mathbf{p}_{Q-1},d_{Q-1})\}$. Here $\mathbf{p}_i$ is the normalized orientation of the plane, and $d_i$ is the distance from the origin; explicitly, these planes are the planes that ``cut'' the region of evaluation into sub-regions. Thus, if we have some point $\mathbf{x} \in R$, we can construct an integer $q$ in the range $\left[0, 2^Q\right)$ by assigning its $i^\text{th}$ bit a value of $1$ if $\mathbf{x}$ is on the left side of the plane $p_i$, and $0$ otherwise. This allows us to associate an integer with every single sub-region of evaluation. However, this does not map to the entire range $\left[0, 2^Q\right)$, that is, it is possible that there are more integers in the range $\left[0, 2^Q\right)$ than there are sub-regions. Further, if $Q$ is large, then $q$ may be vastly bigger than $N$. Therefore, we first compress the range $[0,2^Q)$ down to some range $[0,r)$ --- that is, we first construct $q$ as above, then take its remainder upon division with $r$ (i.e. $q\mod r$) where $r$ is chosen so that each subregion maps to a unique integer in $[0,r)$. To find such an $r$, we simply perform a bruteforce search. We then create another (surjective) map $\sigma: \{0, 1, \cdots, r-1\} \rightarrow \{0, 1, \cdots, N-1\} $ that takes $q \mod r$ to a sub-region's index. This is realized by an array of $r$ entries. On a GPU, we store this in fast constant or shared memory. On a CPU, this will hopefully be cached very quickly. Thus, to determine this index we only need $Q$ plane comparisons, one remainder operation and one fast memory lookup (the conditional behaviour of the plane comparison is done on an instruction predicate level, and does not cause any branching).

For most of the cases we consider, the number of planes $Q$ turns out to be quite small. The largest $Q$ for box splines in 3D we observe is $9$, which corresponds to the quartic BCC spline ~\cite{kim2013quartic}, and seems to have little effect on reconstruction speed. We did however, notice that for all orders of the Voronoi BCC spline, $Q$ is prohibitively large, $Q>32$. While this does not pose a problem to the theory of this method, if we did not impose the compression of the range $[0,2^Q)$ to $[0,r)$ the mapping to sub-regions would consume over $4GB$ of memory. We can additionally apply this procedure recursively. That is, we determine which octant a reconstruction lies within, then transform it to the positive octant; this approach only works provided the spline has the reflective symmetry necessary, which the BCC Voronoi spline does indeed have, and allows us to reduce our lookup tables further. Note that this does introduce some overhead, as one needs to incorporate another transform in the generated code.

\subsection{Branch Free Evaluation via Symmetry Analysis}
\begin{figure}[t!]
    \centering
    \includegraphics[]{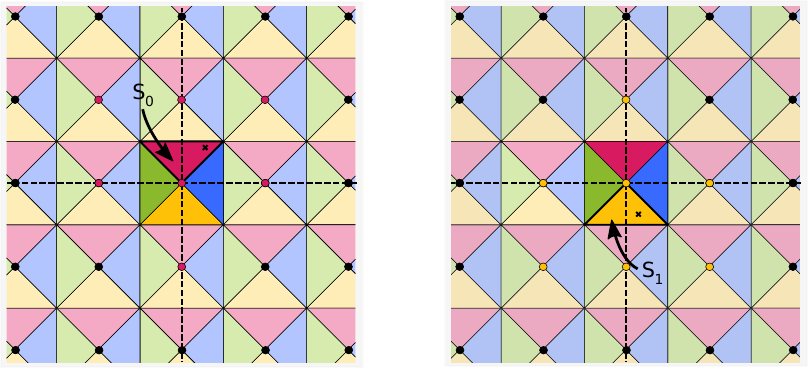}
    \caption{Two different reconstructions within the region of evaluation. Each reconstruction point is denoted by an ``x''. Each reconstructed point lies within a different sub-region of evaluation $S_0$ and $S_1$. The lattice sites that contribute to that reconstruction are coloured with the region's respective color.}
    \label{fig:sub_region_symmetry}
\end{figure}

To build the intuition for this optimization, we begin with an example. Say we have two regions $S_0$ and $S_1$ that ``appear'' similar, Figure~\ref{fig:sub_region_symmetry} shows an example for two sub-regions. Intuitively we can see that $S_1$ is simply a rotation (or a reflection) of $S_0$. One could easily model this with a rotation matrix $T$ and, to be more general, a translation vector $\mathbf{t}$. That is, we would want $ \psi_1(\mathbf{x}, \mathbf{k}) =  \psi_0(T\mathbf{x}-\mathbf{t}, \mathbf{k})$. If this were true, we could store $T$ and $\mathbf{t}$ in a lookup table and implement the piece-wise structure in Equation~\ref{eq:breakdown} as a lookup instead of branch by first looking up $T$ and $\mathbf{t}$, applying them to $\mathbf{x}$, then evaluating a single $\psi$. The situation, however, is slightly more complicated, as we have failed to match the coefficients between $\psi_0$ and $\psi_1$. To make this consistent, we need some ``renaming'' map $\pi$ that renames the $c_\mathbf{n}$ to be consistent with the transformation $T$ and shift $\mathbf{t}$. That is, we desire $T,\mathbf{t},$ and $\pi$ such that  
\begin{eqnarray*}
\psi_0^\pi(T\mathbf{x}-t, \mathbf{k}) & := & \sum_{\mathbf{n} \in C_0(S_0)} c_{\pi(\mathbf{n})+\mathbf{k}}\varphi(T\mathbf{x}-\mathbf{n}- t)  \\
 & = & \sum_{\mathbf{n} \in C_0(S_1)} c_{\mathbf{n}+\mathbf{k}}\varphi(\mathbf{x}-\mathbf{n}) \\ 
 & = & \psi_1(\mathbf{x}, \mathbf{k}).
\end{eqnarray*}
If we find such $T, \mathbf{t}$ and $\pi$ then we know for any $\mathbf{x} \in S_1$ then we have $T^{-1}(\mathbf{x}+\mathbf{t}) \in S_0$, and by renaming the memory look-ups, we can use $\psi_0^{\pi^{-1}}(T^{-1}(\mathbf{x}+\mathbf{t}), \mathbf{k})$ to evaluate the sub-region's function. Here, $\pi$ must be an invertible map, and $T$ must be an invertible rigid body transformation. 

To find such parameters, we first choose a region $S_0$ as our reference sub-region and search for $T, \mathbf{t}$ and $\pi$ for every other sub-region $S_i$. The shift $\mathbf{t}$ is chosen to be the zero vector, or the center of mass for the sub-region; $T$ is chosen from the symmetry group of the spline. We implement this as a combinatorial brute force search. That is, we enumerate all possible pairs of $\mathbf{t}$ and $T$ and check whether this combination yields a valid transformation. Here, ``valid'' means that there exists a $\pi$ such that the above simplification holds. For splines with polynomial sub-regions, we can associate a polynomial with each coefficient lookup $c_\mathbf{m}$ in $\psi_1$, then we apply the transformation to $\psi_2$ and determine the polynomial associated with each $c_\mathbf{m}$ of the transformed sub-region evaluation function. This forms a bipartite graph where the untransformed coefficients are colored in, say blue, and the transformed coefficients are coloured in red; we assign an edge between two points in those sets if their associated polynomial is identical. If there exists a perfect matching between two sets, then we say it is ``valid''. Moreover, the same perfect matching yields the function $\pi$. Perhaps unsurprisingly, $\pi$ is a rigid body transformation in all the cases we consider in this work --- that is, $\pi(\mathbf{n}) = T\mathbf{n} + \mathbf{t}$.

In general, there may be instances where it is not possible to find  $T, \mathbf{t}$ and $\pi$ for two sub-regions $S_i, S_j, i\ne j$. This is the case for the 6 direction box spline on the FCC lattice --- there are sub-regions with vastly different geometry~\cite{kim2013gpu} . In general, we may have $K$ such unique regions that are not re-writeable in terms of one another, and we collect them in the set $\{\psi_0^\pi, \psi_1^\pi, \cdots, \psi_{K-1}^\pi\}$. To handle these cases, we use branch predication --- we calculate the polynomial within all sub-regions and use a predicate operator to choose the correct region's result at the end of the calculation. This performs unnecessary computation, but avoids the heavy overhead of branching on the GPU~\cite{kim2013efficient}. Algorithm~\ref{alg:eval} summarizes how to use the techniques we have discussed to perform point-wise evaluation.

\begin{algorithm}
    \caption{Branch free evaluation at a point.}
    \label{alg:eval}
    \begin{algorithmic}[1] 
        \Procedure{Eval}{$\mathbf{x}$}
            \State $f \gets 0$
            \For{$\mathbf{l} \in \{\mathbf{l}_0,\mathbf{l}_1,\cdots \mathbf{l}_{M-1}\} $} 
            \State $\mathbf{k} \gets \rho(\mathbf{x}-\mathbf{l})$  \Comment{Determine the shift to ROE}
            \State $\mathbf{x}^\prime \gets \mathbf{x} - \mathbf{k}-\mathbf{l}$   \Comment{Shift ROE}
            \State $q \gets 0$
    
            \For{$i \in \{0,1,\cdots Q-1\} $} \Comment{Determine BSP index}
               \State $q \gets (\mathbf{x}^\prime\cdot\mathbf{p}_i - d_i < 0) \ ? \ q : q \mathbin{|} 2^i$
            \EndFor
            \State $SubRegionIndex \gets \sigma(q \% r)$\Comment{Determine the shift to ROE}
            
            \State $T^\prime \gets T[SubRegionIndex]$
            \State $\mathbf{t}^\prime \gets -T^\prime\mathbf{t}[SubRegionIndex]$
            \State $\pi^\prime \gets \pi[SubRegionIndex]$
                
            \State $g \gets 0$
            \For{$i \in \{0,1,\cdots, K-1\} $} 
                \State $v \gets {\psi_i^{\pi^\prime}}(T^\prime\mathbf{x}^\prime - \mathbf{t},  \mathbf{k}+\mathbf{l}) $
                \State $g \gets PsiIndex[SubRegionIndex] == i \ ? \  v \  : \  0$
            \EndFor
            \State $f \gets f + g$\Comment{Add the contribution for this coset}
            \EndFor
            \State \textbf{return} $f$ 
        \EndProcedure
    \end{algorithmic}
\end{algorithm}

\subsection{Reducing Bandwidth Requirements}
\begin{figure}[t!]
    \centering
    \includegraphics[]{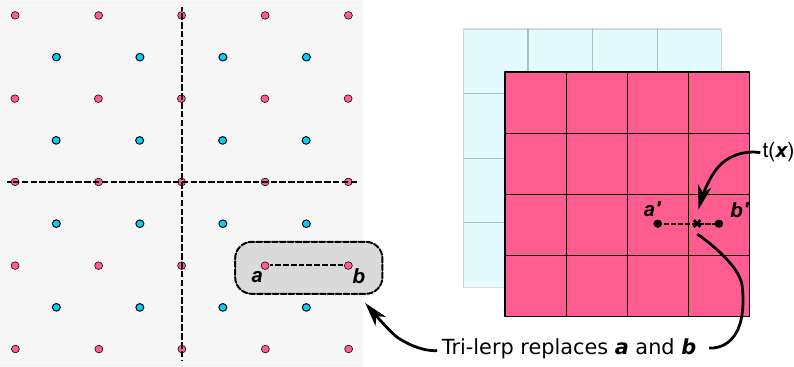}
    \caption{Replacing two texture fetches $\mathbf{a}$ and $\mathbf{b}$ on the quincunx lattice with a single linear texture fetch. There is a small amount of arithmetic involved in translating $\mathbf{a}$ to $\mathbf{a}^\prime$ but it is neglible. Additionally, textels are often shifted slightly to make the hardware design easier --- they are often shifted by 0.5 in the cardinal directions.}
    \label{fig:linear_trick}
\end{figure}

While the above form is appropriate for execution on the GPU, each call to $\psi_i^\pi$ incurs a heavy amount of memory accesses for larger splines. All contemporary GPUs include hardware accelerated linear texture fetches. A naive implementation of the previous sections would use nearest neighbor interpolation to facilitate the coeffecient fetches, however, trilinear texture fetches are only a small constant factor slower (approximately $1.4\times$ on our hardware) and combine up to 8 texture fetches in one instruction.

In general, it is not always possible to use trilinear interpolation to reduce 8 texture fetches into 1 --- this is only possible when the basis is separable. However, it is possible to rewrite two memory fetches (that reside on adjacent points of a given Cartesian coset) as a single tri-linear fetch. This offloads some of the reconstruction bandwidth pressure, but may require more computation. Fortunately, on GPUs, computation is relatively cheap compared to bandwidth.

To demonstrate this, let us fix a single $\psi_i^\pi$, and suppose that we have two two lattice sites $\mathbf{a}$ and $\mathbf{b}$ that are adjacent on the same Cartesian coset. Then we attempt to find two functions $t(\mathbf{x})$, and $g(\mathbf{x})$ such that 
\begin{equation}\label{eq:ltrix}
g(\mathbf{x})\cdot((1-t(\mathbf{x}))c_{\mathbf{a}} + t(\mathbf{x})c_{\mathbf{b}}) = c_{\mathbf{a}} \varphi(\mathbf{x}-\mathbf{a}) + c_{\mathbf{b}}\varphi(\mathbf{x}-\mathbf{b}).
    \end{equation}
The term $((1-t(\mathbf{x}))c_{\mathbf{a}} + t(\mathbf{x})c_{\mathbf{b}})$ is exactly linear interpolation; this can be easily translated into single texture fetch instruction (and some additional computation for $t(\mathbf{x})$ and $g(\mathbf{x})).$ Explicitly, the solutions $g(\mathbf{x})$ and $t(\mathbf{x})$ are given by
$$
g(\mathbf{x}) = \varphi(\mathbf{x}- \mathbf{a}) +\varphi(\mathbf{x}- \mathbf{b})  \ \text{ and  } \ t(\mathbf{x}) = \frac{\varphi(\mathbf{x}- \mathbf{b}) }{g(\mathbf{x})}
$$
which is easily verified. Note that we require $|g(\mathbf{x})| > 0$ for this to be valid --- to see why this is true in our case, we fix $\mathbf{x}$ within a given sub-region. Since we wish to use Equation~\eqref{eq:ltrix} to replace two fetches in the convolution sum~\eqref{eq:conv_sum}, we know that the basis shifts associated with two coefficients in that sum must contribute to the final sum over the entire sub-region --- they cannot be identically zero. The only other troublesome case is if $\varphi(\mathbf{x}-\mathbf{a}) = -\varphi(\mathbf{x}-\mathbf{b})$ for some $\mathbf{x}$ in the sub-region under investigation. For all the cases we consider, our basis functions are strictly positive, so we need not worry about this --- however, for some interpolating bases, this may present a problem. If we are working with volumetric data, with $\mathbf{a}^\prime$ and $\mathbf{b}^\prime$ defined as the texture coordinates of the lattice sites $\mathbf{a}$ and $\mathbf{b}$ in GPU memory, then we can write Equation~\ref{eq:ltrix} as  
$$
g(\mathbf{x})\cdot TEX3D(\text{tex\_coset}, \mathbf{a}^\prime + t(\mathbf{x})(\mathbf{b}^\prime - \mathbf{a}^\prime ))
$$
where $TEX3D$ is the linear texture fetch built into most GPUs. This is demonstrated in Figure~\ref{fig:linear_trick}. It also possible to perform a similar optimization with groups of 4 points that are arranged as a square (on a Cartesian coset) and groups of 8 arranged as a cube (again, on a Cartesian coset). Algebraically, these cases are much more involved, but the basic principle of solving for two functions $t(\mathbf{x})$ and $g(\mathbf{x})$ remains the same.

While we may be able to re-write groups of fetches as a single linear fetch, there are some details that need to be addressed. The first subtlety is that, while it may be possible to cover two memory fetches with one, it may also be possible to group those two memory fetches in a larger group. If every grouped memory fetch is a subset of the total collection of memory fetches, then we seek a minimal {\em set covering} of the total set.

\begin{figure}[t!]
    \centering
    \includegraphics[scale=1.5]{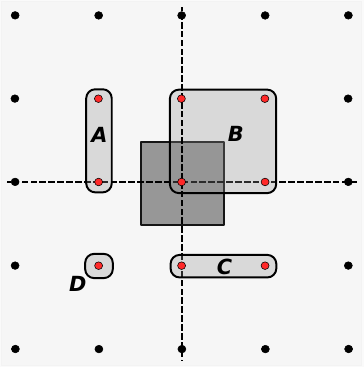}
    \caption{The set covering for the tensor product quadratic spline. The middle square is the region of evaluation (there is only one sub-region of evaluation for this spline). The points in red denote the lattice sites that contribute to a reconstruction in the region of evaluation. The highlighted and labeled boxes correspond to the groups of 1, 2, and 4 that can be simplified into a linear texture fetch. Note that, by our heuristic, we would rather evaluate these groups in the order D-C-B-A as opposed to, say, D-B-C-A which would have $4+2+2+2=10$ and $4+3+2+3=12$ cache misses, respectively.}
    \label{fig:grouping}
\end{figure}
The set covering problem is NP-complete, however, we only perform this step once as a precomputation. We enumerate all groups of 1, 2, 4 and 8 that can be combined into a single texture fetch and use Algorithm X to find a solution to the minimal set covering problem~\cite{knuth2000dancing}. Once we have a set covering we may still choose an ordering of the memory fetches. We make the assumption that, when a texture fetch occurs, all the points that contribute to that texture fetch are cached --- any neighboring texture fetch may then reuse some of the values that have been cached. To take advantage of this, we create a complete directed-graph in which each node corresponds to one linear texture fetch. To each edge of the graph we assign the number of cache-misses incurred by performing the pair of memory fetches (in order). We then seek a route through this graph that minimizes cache misses --- this reduces to the travelling salesman problem; another NP-complete problem. Again, this is a one-time precomputation, and Figure~\ref{fig:grouping} shows an example of the set covering for the quadratic tensor product spline.

\section{Experiments \& Results} \label{sec:results}
In this work, we validate our methodology by demonstrating that our framework works for many different test cases. While performance is important, and we do report limited performance metrics in this work, we defer in-depth performance analysis to Part II~\cite{part2}. To this end, we choose two simple applications that demonstrate the generality of our framework. The first is volumetric rendering, in which we generate implementations for various different spline lattice combinations, many of which have not had fast GPU implementations. The second application is simply function approximation, however, we do this on the $\mathcal{D}_4$ lattice on the CPU. To our knowledge, there relatively little works that investigate function approximation in this space. We are not aware of any that assess the convergence of such spaces, nor any that provide implementations.

We do not ``implement'' Algorithm~\ref{alg:eval} directly; we use it simply as a template for code generation. That is, the input to our pipeline is a list of (polyhedral region, polynomial) pairs, and the output is generated LLVM code~\cite{LLVM}. For our volumetric rendering experiments, we use the LLVM code to generate PTX code, for our other tests we emit x86\_64 assembly. We have constrained ourselves to the CUDA ecosystem as it is the most popular GPGPU platform, however, LLVM contains both a backend for AMD GPUs and a platform agnostic SPIR-V backend that is capable of targing OpenCL + OpenGL devices~\cite{kessenich2018spir}. The details of the code generation stage, and a more detailed breakdown of performance is detailed in part II~\cite{part2}.

The processing time for a given interpolant varies based on the spline, higher order splines with large support take much longer, but the process takes typically on the order of hours on an Intel i7 7700 with 16GB of DDR4 RAM at 2333 MHz. However, the processing time of the pipeline is a one time computation, and need not be repeated once an effecient form has been derived. We run our tests on the generated code.

\subsection{Volume Rendering}
Volume rendering is an illustrative problem as it demonstrates an average case application --- as rays march through the cells in the volume, encountering a new cell will cause a cache miss, however, for points sampled within a cell we get cache hits. Moreover it does not favour any given lattice structure --- a problem that uses slices of a volume along principle lattice directions may favour one case over another.

All of our tests were run with the same CPU as above and an NVIDIA GTX 1060 with 6GB ram and no memory or core over-clocks. Compute kernels were set to use a maximum of 128 registers. Table~\ref{tab:interpolants} enumerates the list of interpolants for which we generated code. When generating code, there is a small number of parameters to choose --- we tuned each interpolant separately in order to maximize their performance as detailed in Part II~\cite{part2}. 

Each test cased was budgeted approximately the same amount of memory to ensure fairness between test results. For our test function, we sampled the Marshner Lobb function~\cite{marschner1994evaluation} at each lattice site, this is the de-facto standard when comparing interpolation kernels --- as one moves further from the center of the volume, the spatial frequency increases, as such reconstruction errors visually depict how frequency content is distorted. Typically, one renders the Marshner Lobb as an iso-surface, however we stick to volumetric rendering to expose the worst case behaviour of the algorithm. The isosurface for a given isovalue may change slightly among grids, thus one grid may terminate earlier than others, our volumetric rendering scheme ensures uniformity of marched rays. Scale was chosen to be equal in each case. Our transfer function was chosen so that the volume was mostly transparent, thus forcing all rays to traverse the entire volume. 

\setlength{\tabcolsep}{1pt}
\renewcommand{\arraystretch}{1.3}

\begin{table}

\begin{tabular}{lcllccc}
 Interpolant Name & Lattice & \multicolumn{2}{c}{Direction Matrix} & Order & \# Lookups  & Reference  \\    \hline\hline
 \makecell{CC Voronoi Spline 1} & CC & \multicolumn{2}{c}{\scalemath{0.75}{\cctpbsl^{:2}}} & 2 & 8 & \ \\
 \makecell{{FCC Voronoi Spline 1}$^\dagger$} & CC & \multicolumn{2}{c}{N\textbackslash{A}} & 2 & 16 & \makecell{\cite{mirzargar2010voronoi}\\ \cite{mirzargar2011quasi}} \\ 
 \makecell{Linear Rhombic$^\dagger$ \\ Dodecahedron} & CC & \multicolumn{2}{c}{\scalemath{0.75}{\bccrodecbs}} & 2 & 16 &\makecell{\cite{entezari2004linear} \\ \cite{retailor}} \\
 \makecell{CC Voronoi Spline 2} & CC & \multicolumn{2}{c}{\scalemath{0.75}{\cctpbsl^{:3}}} & 3 & 27 & \ \\
 \makecell{BCC Voronoi Spline 1$^\dagger$} & CC & \multicolumn{2}{c}{N\textbackslash{A}} & 2 & 32 & \makecell{\cite{mirzargar2010voronoi}\\ \cite{mirzargar2011quasi}} \\
 \makecell{Truncated Rhombic \\ Dodecahedron} & CC & \multicolumn{2}{c}{\scalemath{0.75}{\cczpbs}} & 4 & 53 & \cite{entezari2006extensions}  \\
 \makecell{FCC Voronoi Spline 2$^\dagger$} & CC & \multicolumn{2}{c}{N\textbackslash{A}} & 3 & 54 & \makecell{\cite{mirzargar2010voronoi} \\ \cite{mirzargar2011quasi}}  \\
 \makecell{CC Voronoi Spline 3} & CC & \multicolumn{2}{c}{\scalemath{0.75}{\cctpbsl^{:4}}} & 4 & 64 & \ \\
 \makecell{Linear Rhombic \\ Dodecahedron} & BCC  &\multicolumn{2}{c}{\scalemath{0.75}{\bccrodecbs^{\phantom{:1}}}}   &  2 & 4 & \cite{entezari2004linear} \\
 \makecell{BCC Voronoi Spline 1$^\dagger$} & BCC & \multicolumn{2}{c}{N\textbackslash{A}} & 2 & 8 & \makecell{\cite{mirzargar2010voronoi}\\ \cite{mirzargar2011quasi}} \\
 \makecell{BCC Voronoi Spline 2$^\dagger$} & BCC & \multicolumn{2}{c}{N\textbackslash{A}} & 3 & 27 & \makecell{\cite{mirzargar2010voronoi} \\ \cite{mirzargar2011quasi}} \\
  \makecell{Quartic Truncated \\ Rhombic Dodecahedron} & BCC  &\multicolumn{2}{c}{\scalemath{0.75}{\bcczpbs}}   &  4 & 30 & \cite{kim2013quartic} \\
\makecell{Qunitic Rhombic \\ Dodecahedron} & BCC  &\multicolumn{2}{c}{\scalemath{0.75}{\bccrodecbs^{:2}}}   &  4 & 32 & \cite{entezari2004linear} \\
 \makecell{FCC Voronoi Spline 1$^\dagger$} & FCC & \multicolumn{2}{c}{N\textbackslash{A}} & 2 & 8 & \makecell{\cite{mirzargar2010voronoi} \\ \cite{mirzargar2011quasi}}  \\
\makecell{Cubic Truncated \\ Octohedron} & FCC  &\multicolumn{2}{c}{\scalemath{0.75}{\fccsix}}   &  3 & 16 & \cite{kim2008box} \\
 \makecell{FCC Voronoi Spline 2$^\dagger$} & FCC & \multicolumn{2}{c}{N\textbackslash{A}} & 3 & 27 & \makecell{\cite{mirzargar2010voronoi}\\ \cite{mirzargar2011quasi}}  \\
\end{tabular}
\caption {\label{tab:interpolants} A list of all interpolants and lattices for which we generated code. The notation $:n$ denotes concatenating a matrix with itself $n -1$ times. Splines are named by the shape of their support, and the degree of the polynomial pieces, except for the Voronoi splines. Splines that cannot be generated by a single box spline have N\textbackslash{A} in the Direction Matrix column. Splines marked with a $\dagger$ are ones for which there exists no previous GPU implementation.} 
\end{table}

\subsubsection{Quantitative Results}
\begin{figure}
    \centering
    \includegraphics[width=0.8\textwidth]{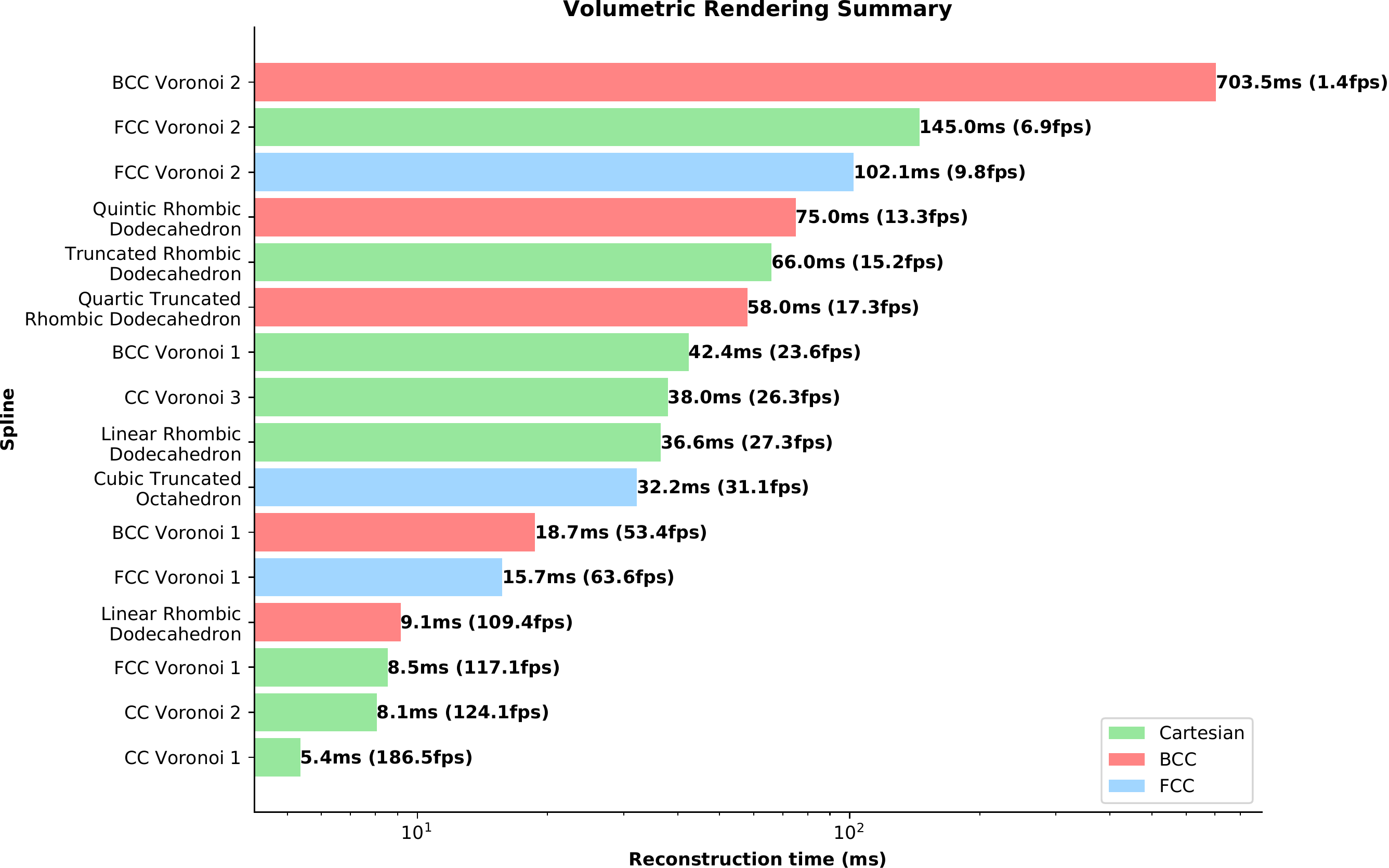}
    \caption{Reconstruction speed in milliseconds per frame (alternatively FPS) for each test case.}
    \label{fig:results1}
\end{figure}

For every combination of basis function and lattice tested, we obtained reasonable render times, shown in Figure~\ref{fig:results1}. The only case in which performance degraded beyond what we consider reasonable is the BCC Voronoi 2 spline on the BCC lattice --- this is likely due to the complex polynomial structure of the spline combined with the large amount of memory fetches one needs to reconstruct a single value. We do not consider this spline to be real time, however it may be used in progressive rendering approaches to refine a rendered image once user interaction has stopped. 

\begin{figure}
    \centering
    \includegraphics[width=0.5\textwidth]{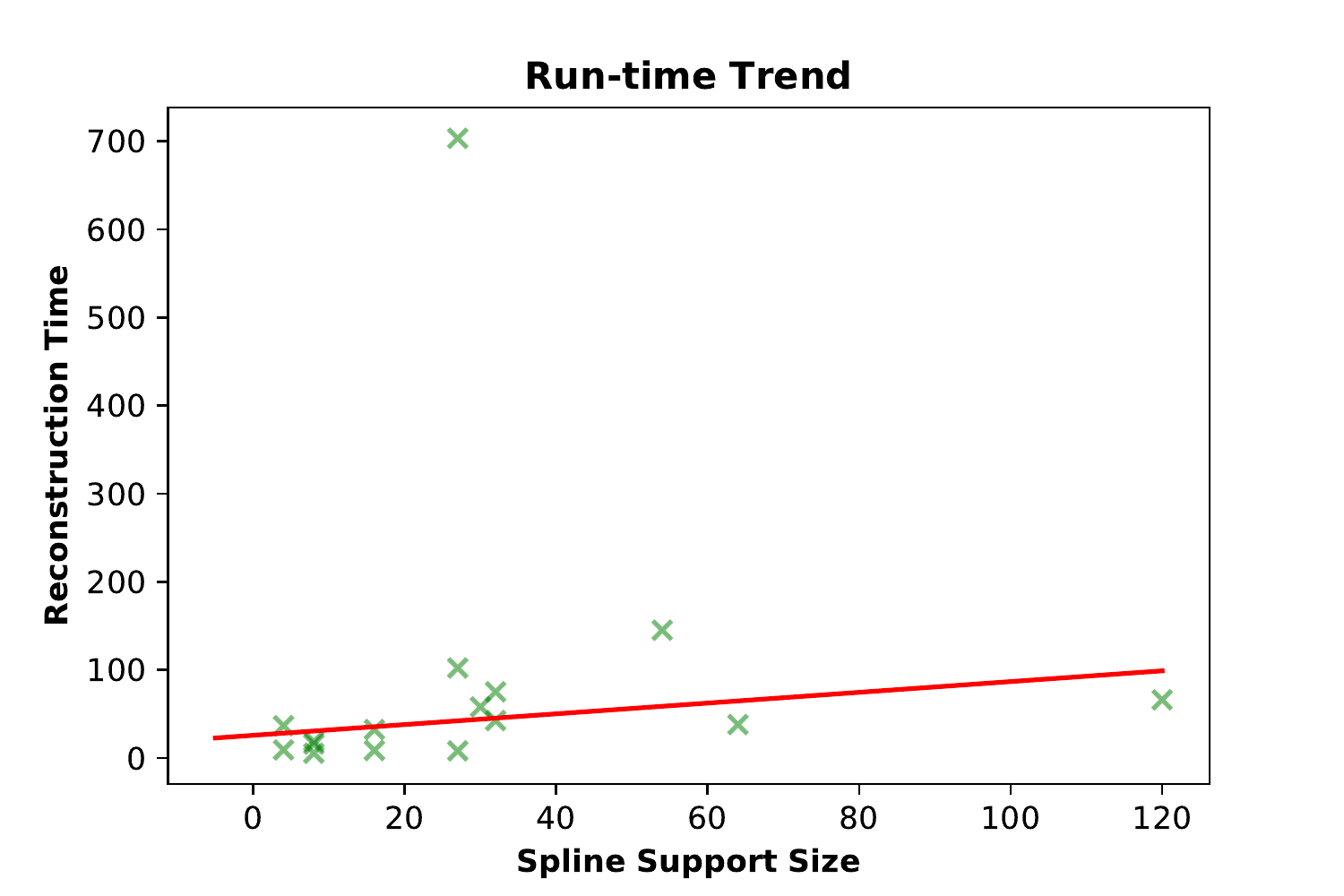}
    \caption{Trend of reconstruction speeds. The red line is the least squares fit without the outlier at (27, 700ms) --- this corresponds to the BCC Voronoi 2, which is a very computationally expensive spline on the GPU, it requires the predication of unique sub-regions of evaluation.}
    \label{fig:results2}
\end{figure}

A question one might have, based on these results, is how the generated code scales up as the interpolants become more complex. That is, if we require double the memory accesses per reconstruction, it is reasonable to think that a reconstruction would take double the amount of time to complete. Figure~\ref{fig:results2} shows a plot relating the number of points that contribute to a reconstructed point versus the render time, and a least squares fit (with the single outlier point removed). The least squares fit has a slope of approximately 0.6. At first glance this appears quite promising --- one might expect a slope of 1, and 0.6 implies that the cost of doubling the complexity of the spline is less than one might expect. However, the correlation coefficient of these data is approximately 0.45 (with p-value 0.07), thus this correlation is quite weak. Unfortunately, there are too many confounding variables --- remember, we take the best results over all configurations of spline spaces (i.e we mix branch predication, tri-linear lookups, etc.). We will return to this in Part II.

\subsubsection{Qualitative Results}
\begin{figure}[ht!]
  \subfloat[Ground Truth]{
	\begin{minipage}[c][1\width]{
	   0.38\textwidth}
	   \centering
	   \includegraphics[width=1\textwidth]{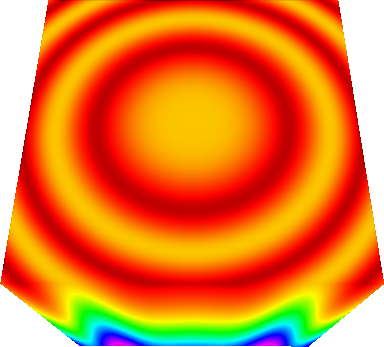}
	\end{minipage}}
  \subfloat[CC Voronoi 1]{
	\begin{minipage}[c][1\width]{
	   0.38\textwidth}
	   \centering
	   \includegraphics[width=1\textwidth]{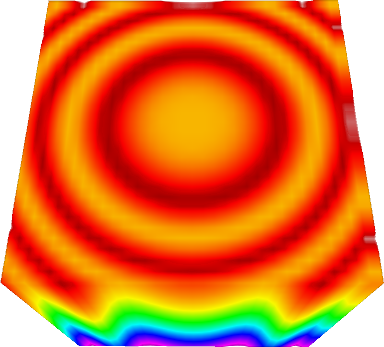}
	\end{minipage}}
 \hfill 	\\
  \subfloat[BCC Voronoi 1]{
	\begin{minipage}[c][1\width]{
	   0.38\textwidth}
	   \centering
	   \includegraphics[width=1.0\textwidth]{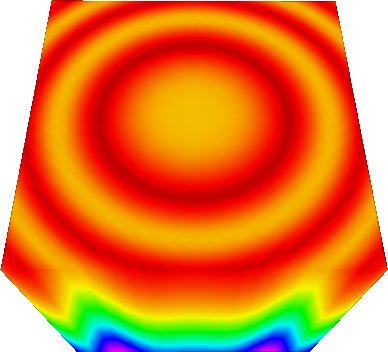}
	\end{minipage}}
  \subfloat[FCC Voronoi 1]{
	\begin{minipage}[c][1\width]{
	   0.38\textwidth}
	   \centering
	   \includegraphics[width=1.05\textwidth]{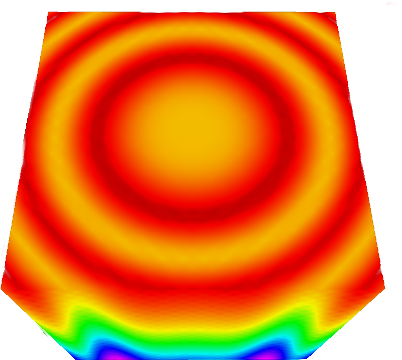}
	\end{minipage}}
\caption{Voronoi spline reconstructions on different lattices. Note the much harsher artifacting on the top of the volume for the CC lattice. The FCC and BCC lattices show much smoother reconstructions, with the BCC having slightly less isotropic reconstruction on the top, and the FCC having slightly less isotropic reconstruction on the lower face. These were all rendered with framerates well above 60fps.} \label{fig:qualitative}
\end{figure}

While it is not the goal of this work to showcase qualitative visual results, we compare the $V_1$ splines of each lattice on their respective lattice. Figure~\ref{fig:qualitative} shows the qualitative difference for the rendered volumes. These are at similar resolutions, and it is clear that the BCC and FCC outperform the CC lattice. Between the BCC and FCC lattice the comparison is more difficult, there are subtle difference in reconstruction, yet the artifacting seems wholly more isotropic on the BCC lattice. 

\subsection{Approximation on the $\mathcal{D}_4$ lattice}
To demonstrate that we are not bound to 2 and 3 dimensions, we derive a fast evaluation scheme for the 4-dimensional $\mathcal{D}_4$ lattice. The generating lattice, as well as the direction matrix for our spline are
\begin{equation}
    L_{\mathcal{D}_4}:=\begin{bmatrix}
-1 & \phantom{-}0 & \phantom{-}0 & \phantom{-}0 \\
\phantom{-}1 & \phantom{-}0 & -1 & -1 \\
\phantom{-}0 & -1 & \phantom{-}0 & \phantom{-}1 \\
\phantom{-}0 & -1 & \phantom{-}1 & \phantom{-}0
    \end{bmatrix}
    \;\; \text{and} \;\;
    \Xi_{\mathcal{D}_4}:=\begin{bmatrix}
\phantom{-}1 & \phantom{-}1 & \phantom{-}1 & \phantom{-}1 & -1 & -1 & -1 & -1 \\
\phantom{-}1 & \phantom{-}1 & -1 & -1 & \phantom{-}1 & \phantom{-}1 & -1 & -1 \\
\phantom{-}1 & -1 & 1 & -1 & \phantom{-}1 & -1 & \phantom{-}1 & -1 \\
\phantom{-}1 & \phantom{-}1 & \phantom{-}1 & \phantom{-}1 & \phantom{-}1 & \phantom{-}1 & \phantom{-}1 & \phantom{-}1
    \end{bmatrix}
\end{equation}
respectively. The geometry of this lattice is similar to that of the FCC lattice; the $\mathcal{D}_4$ lattice consists of 8 Cartesian cosets, shifted to the 3-facets of the 4-dimensional hypercube. The spline we choose is a modification of the one presented by Kim et al., however some direction vectors have been removed to make the space computable in a reasonable amount of time~\cite{kimroot}. The spline $M_{\Xi_{\mathcal{D}_4}}(\mathbf{x})$ is a fourth order spline, but not interpolating, as such it must be pre-filtered to ensure that the error will decay as expected. We used a filter with value $\frac{7}{3}$ at the origin, and $-\frac{1}{18}$ at all $\mathcal{D}_4$ lattice sites at distance $\sqrt{2}$ from the origin. We approximated a Gaussian with mean $\boldmath{0}$ and $\sigma=0.125$. We started with a grid scale $h=0.5$ and successively halved $h$ 10 times, at each iteration we sampled the function on the grid $h\cdot L_{\mathcal{D}_4}$, then measured the $L^2$ error over the domain $[-0.5, 0.5]^4$ via Monte Carlo integration with 10,000 samples. The overall decomposition analysis for the spline took approximately 2 days to compute, and the code generation~\cite{part2} took approximately a day to compute. Again, this is a one time pre-computation; reconstruction, on the other hand, is many orders of magnitude faster, a single point evaluation takes less than a millisecond. 

\subsubsection{Results}
\begin{figure}[h!]
    \centering
    \includegraphics[width=0.5\textwidth]{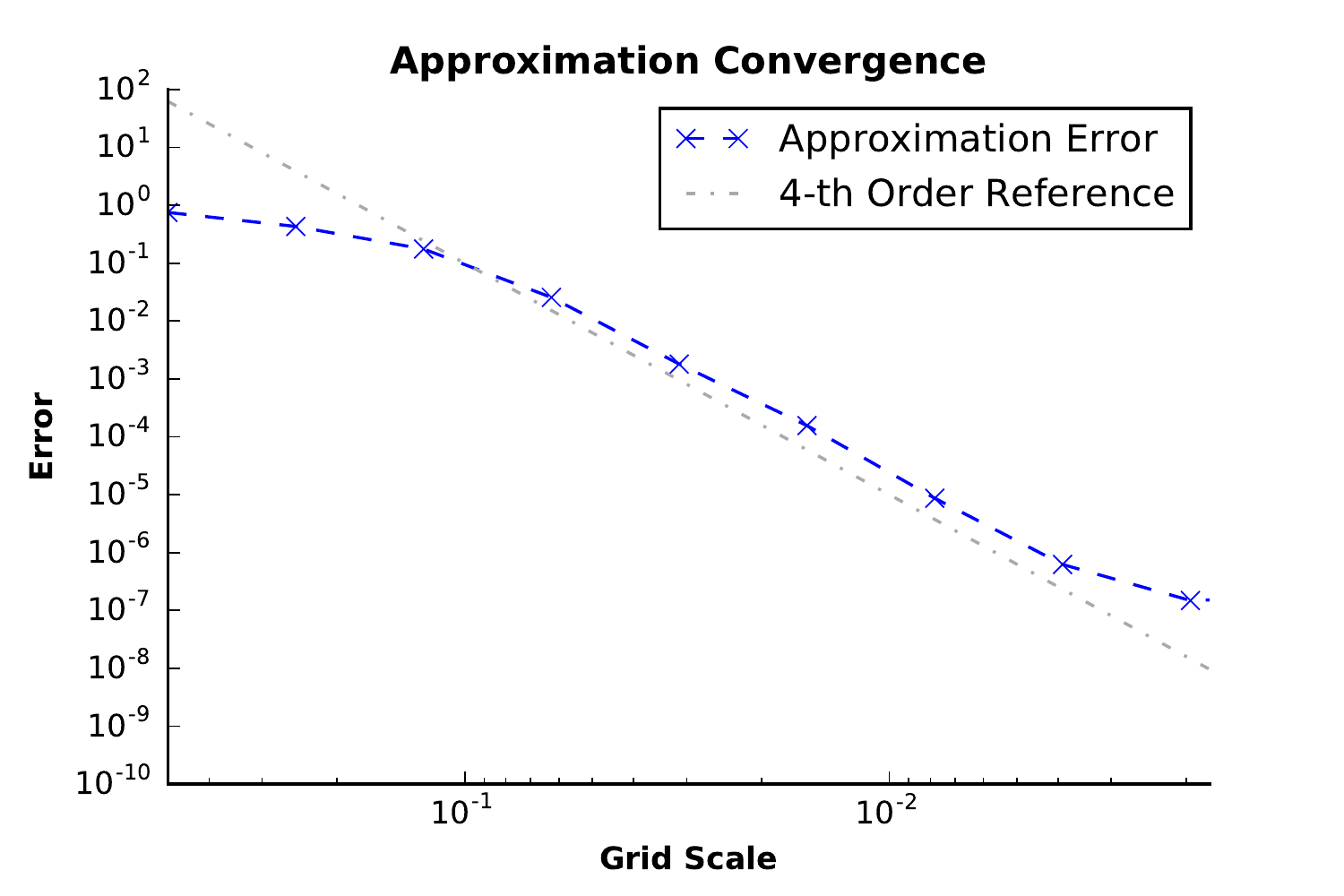}
    \caption{Error decay of  function approximation as the $\mathcal{D}_4$ lattice  becomes finer. The slope of the light grey line indicates the theoretic decay of the approximation space, and the line denotes the actual decay of the approximation error. The function being approximated is a Gaussian. }
    \label{fig:d4conv}
\end{figure}

Since the spline $M_{\Xi_{\mathcal{D}_4}}(\mathbf{x})$ is fourth order spline, we expect to see the error decay by a factor of $\frac{1}{16}$ at every iteration. Figure~\ref{fig:d4conv} demonstrates exactly this behaviour. An equivalent tensor-product spline on the 4-dimensional Cartesian lattice would require 256 memory accesses for the same order, whereas this case requires only 240. Perhaps astonishingly, the same spline has the same order on the $\mathcal{D}^\ast_4$ lattice (the dual of the $\mathcal{D}_4$) requiring only 60 memory accesses. The $\mathcal{D}_4$ lattice attains the optimal hyper-sphere packing in 4-dimensions, as such, the $\mathcal{D}^\ast_4$ lattice would be the optimal sampling lattice. However, we stick to the $\mathcal{D}_4$ lattice, as it is more difficult to derive interpolants for; its coset structure contains more shifted lattices than the $\mathcal{D}^\ast_4$ lattice, and is somewhat of a ``stress-test'' for our methodology.

\section{Conclusion}
We presented a generalized framework for analysing multidimensional splines on non-Cartesian (and Cartesian grids), with the target of producing fast evaluation schemes for said spline spaces. While this is the main contribution of the work, we also produced performant code for the notoriously complex Voronoi splines on the FCC and BCC lattice which have not yet had efficient implementations. We also demonstrated the computational behaviour of our approach as spline size increased, showing reasonable computational increase as the support of a spline increases---however we saw that it is difficult to predict performance based on support size alone. Finally, we investigated the performance in 4-dimensions, and provided an imlpementation (and quasi-interpolant filter) for a spline on the $\mathcal{D}_4$ lattice. The entire pipeline of our worksheet is implemented within a SageMath worksheet, and is available on github~\cite{sage, fastsplinegit}. Further details related to the code generation step of our pipeline, as well as detailed performance results are presented in part II of this work~\cite{part2}. Future theoretical work is focused on extending this framework to more classes of splines (i.e. the exponential box splines), designing optimized interpolants and assessing the quality of interpolants in a systematic manner.


\bibliography{main}
\bibliographystyle{plainnat}
\end{document}